\documentclass{article}
\usepackage{arxiv}
\usepackage{amsmath}
\usepackage[utf8]{inputenc} 
\usepackage[T1]{fontenc}    
\usepackage{hyperref}       
\usepackage{url}            
\usepackage{booktabs}       
\usepackage{amsfonts}       
\usepackage{nicefrac}       
\usepackage{microtype}      
\usepackage{lipsum}
\usepackage{graphicx}
\graphicspath{ {./images/} }
\usepackage[authoryear]{natbib}
\RequirePackage{graphicx}
\usepackage{tikz} 
\usepackage{verbatim}
\usepackage{caption}
\usepackage{subcaption}
\usepackage{bbm}
\usepackage{mathabx}
\usepackage{blindtext}
\usepackage{xcolor}
\usepackage{amsmath,amsthm,amsfonts,amssymb}
\RequirePackage{amsthm,amsmath,amsfonts,amssymb}
\newtheorem{assumption}{Assumption}

\newtheorem{theorem}{Theorem}
\newtheorem{definition}{Definition}
\title{High-Dimensional Covariate-Dependent Discrete Graphical Models and Dynamic Ising Models}

\author{
 Lyndsay Roach \thanks{These authors are co-first authors and contribute equally to this work. $\dagger$ For correspondence, contact Gao at xingao@yorku.ca.}\\
  Department of Mathematics and Statistics\\
  York University\\
  Toronto, M3J 1P3 \\
  \texttt{lyndsay.roach@gmail.com} \\
   \And
 Qiong Li\,$^*$ \\
  Guangdong Provincial/Zhuhai Key Lab of\\ Interdisciplinary Research and Application for Data Science\\
   Beijing Normal-Hong Kong Baptist University\\
  Zhuhai 519087, China \\
  \texttt{qiongli@uic.edu.cn} 
   \AND
    Nanwei Wang\,$^*$ \\
  Department of Mathematics and Statistics\\
  University of New Brunswick\\
  Fredericton, E3B 5A3, Canada \\
  \texttt{nanwei.wang@unb.ca} 
   \And
   Xin Gao $\dagger$\\
   Department of Mathematics and Statistics \\
  York University\\
  Toronto, ON M3J1P3, Canada
  \texttt{xingao@yorku.ca} \\
}

\begin{document}
\maketitle
\begin{abstract}
We propose a covariate-dependent discrete graphical model for capturing dynamic networks among discrete random variables, allowing the dependence structure among vertices to vary with covariates. This discrete dynamic network encompasses the dynamic Ising model as a special case. We formulate a likelihood-based approach for parameter estimation and statistical inference. We achieve efficient parameter estimation in high-dimensional settings through the use of the pseudo-likelihood method. To perform model selection, a birth-and-death Markov chain Monte Carlo algorithm is proposed to explore the model space and select the most suitable model.
\end{abstract}

\thispagestyle{fancy}
\fancyfoot[C]{\small \textcopyright~2025 Xin Gao.
 This work is licensed for academic and non-commercial use only. 
}

\section{Introduction}
Graphical models are essential for understanding complex dependencies among variables, especially in high-dimensional settings. These models represent variables as nodes and conditional independencies as edges, drawing on classical results from graph theory to model probabilistic relationships. See \citet{lauritzen1996gms} for foundational concepts. Graphical models can accomodate continuous, discrete, or mixed variable types, and they can be represented by either directed or undirected graphs. In directed graphical models, edges denote causal dependencies between variables, while in undirected models, edges denote associations without a specific direction of influence.

By bridging the gap between graph theory and probability theory, graphical models provide a structured framework for representing and interpreting complex distributions, laying the groundwork for efficient computation in high-dimensional settings. As discussed in \citet{wainwrightjordan2008}, the connection between graphical models and exponential family distributions demonstrates how these models can represent the factorization of joint distributions, with the factors corresponding to exponential family distributions. Such factorization allows for efficient handling of large-scale probabilistic systems. Moreover, graphical models naturally align with Bayesian statistical analysis. The introduction of hyper Markov laws, by \citet{lauritzendawid1993}, ensures that prior and posterior distributions remain consistent with the graphical structure, further enabling a broad range of inferential methods to be employed. 

Graphical models facilitate the efficient processing and analysis of large datasets, making them a fundamental tool for identifying underlying patterns and predicting outcomes across various fields such as genetics, finance, and social network analysis. As a result, a growing body of research is focused on exploring their advantages. For example, \cite{roverato2002hyper} developed the hyper inverse Wishart distribution for non-decomposable graphs, along with further work on Gaussian graphical models (GGMs) constructed using the DY-conjugate prior within the Bayesian framework. \cite{castelo2006robust} provided a robust method for model selection in Gaussian graphical models when the dimensionality $p$ exceeds the sample size $n$. The authors apply this method to learning large-scale biomolecular networks.  More recently, \cite{model_selection_2022} introduced a Bayesian model selection framework for permutation-invariant Gaussian models, specifically examining random variables invariant under subgroups of the symmetric group. \citet{RoachMassamWangGao2025} demonstrate strong model selection consistency for discrete graphical model when $p$ increases with sample size by articulating the asymptotic behavior of the Bayes factor with hyper Dirichlet prior, as established by \citet{lauritzendawid1993}. In addition, \citet{RoachGao2023} propose a local genetic algorithm with a crossover-hill-climbing operator for discrete graphical models that exploits the relationship between non-decomposable graphs and minimal triangulations.

Here, we focus on discrete graphical models, which are designed to analyze data organized in contingency tables, where $n$ objects are classified according to $p$ criteria. Consider a vector of random variables $\boldsymbol{Y}=(Y_{v}, v\in V)$ indexed by the set $V=\{1, 2, \ldots, p\}$ such that each $Y_{v}$ takes values in a $p$-dimensional contingency table. The conditional independencies between the random variables $Y_{v}$ can be inferred from an undirected graph $G=(V,E)$ with vertex set $V$ and edge set $E\subseteq V\times V$. The discrete graphical model for $\boldsymbol{Y}$ is said to be decomposable or Markov with respect to $G$, if $Y_{a}$ is independent of $Y_{b}$ given $Y_{V\backslash \{a,b\}}$, whenever $(a,b)$ is not an edge in $E$. It is assumed that the cell counts of the contingency table follow a multinomial distribution, with the cell probabilities modeled by a hierarchical log-linear model. The class of discrete graphical models that are Markov with respect to an undirected graph $G$ is a subclass of the class of hierarchical log-linear models, as described in \citet{2012Bayes},  \citet{massamliudobra2009} and \cite{kolodziejek2023discrete}. 

In the subsequent sections, we introduce a covariate-dependent log-linear model that extends traditional discrete graphical models to enable them to accommodate dynamic systems. These dynamic discrete graphical models offer a robust framework for understanding and managing systems with evolving dependencies. They are especially valuable in scenarios where temporal factors impact system behaviour, providing powerful tools for prediction, adaptation, and decision-making. For example, in healthcare, such models can monitor disease progression by incorporating patient-specific covariates, while in finance, they aid in understanding market behaviors and assessing risk factors over time. 

More recently, significant advancements have been made in incorporating covariates into GGMs. For instance, \citet{guo2011joint} and \citet{lin2017joint} introduced multiple GGMs to estimate common and distinct structures across different discretized groups. \citet{liu2010graph} first clustered the covariates into small spaces based on a regression tree and then estimated the GGMs in each subspace separately. \citet{zhang2023high} provided Gaussian graphical regression models that characterize covariate-dependent network structures, accommodating both discrete and continuous covariates. This innovative approach enables the linear adjustment of both the mean and the precision matrix in response to the covariates.

In the Bayesian framework, various approaches have been proposed to estimate heterogeneous graphical models for group-specific graphs \citep{peterson2015bayesian,Tan_2017,10.1214/14-BA931,lin2017joint}. These methods are motivated by progress made in the joint estimation of multiple graphs. \citet{ni2022bayesian} introduced Bayesian Gaussian graphical models with covariates, incorporating additional covariates into the estimation process. This model uses a mixed prior for precision matrices, encompassing a wide range of existing graphical models for heterogeneous settings. Furthermore, to avoid directly parameterizing the precision matrix in the graphical model, \cite{niu2024covariate} proposed a novel covariate-dependent Gaussian graphical model based on random partition. They employed a conjugate \( G \)-Wishart prior on the Gaussian likelihood distribution and use a product of Gaussian conditionals in the pseudo-likelihood approach to investigate the conditional independence within the graph. Covariate-dependent Gaussian graphical models are effectively used to analyze a variety of multi-omics and spatial transcriptomics datasets, aiming to explore inter- and intra-sample molecular networks, as detailed in \cite{CHEN2025100984}. For discrete graphical model,
\cite{cheng2014sparse} proposed a sparse covariate dependent Ising model to study both the conditional dependency within the binary data and its relationship with the additional covariates.

In this article, we present a dynamic discrete graphical model with a novel covariate-dependent log-linear parametrization. To our knowledge, this approach has not been explored in existing literature, which primarily focuses on covariate-dependent Gaussian graphical models. Our method extends the classical log-linear model parameter by incorporating a baseline vector and a slope vector which has associated covariates. Following the notation established by \citet{2012Bayes}, we derive and present the log-likelihood function, score vector, and Hessian matrix for our covariate-dependent log-linear model. We leverage the likelihood framework to establish asymptotic normality of the maximum likelihood estimate (MLE) and formulate various hypothesis tests. We use the well-known Ising model to exhibit our dynamic discrete graphical model in a high-dimensional setting and explain how to use the pseudo-likelihood to perform parameter estimation when the Ising model structure is known. To validate our theoretical results, we conduct comprehensive simulation studies to showcase the performance of the proposed likelihood procedures and pseudo-likelihood method for parameter estimation of a high-dimensional Ising model. In terms of dynamic Ising model structure learning problems, we adopt scablable birth-death MCMC (SBDMCMC) algorithm developed in \citet{mohammadi2015bayesian}, \citet{dobra2018loglinear} and \citet{wang2023scalable}. 
In addition, we illustrate the practical application of our model with a dataset of gene expression related to influenza vaccination, showcasing its utility in real-world scenarios.

The remainder of this paper is organized as follows: In Section \ref{sec:2}, we introduce our model setup and notation. Then, we present theoretical results and statistical inferences. Section \ref{sec:3} delves into the dynamic Ising model, applying the pseudo-likelihood method for parameter estimation in a high-dimensional setting. In Section \ref{sec:4}, we propose Bayesian model selection method for covariate-dependent discrete graphical
models. We substantiate our proposed model and theoretical findings through both small-scale and large-scale numerical simulations in Section \ref{sec:5}, and apply our methods to a real dataset in Section \ref{sec:6}, analyzing influenza vaccination data.

\section{Model Setup and Statistical Inference}\label{sec:2}

In this section, we first present the classical setup for a static discrete graphical model. We then introduce our covariate-dependent discrete graphical model and provide the corresponding theoretical results.

\subsection{Static discrete graphical model }\label{sec:2.1}
Consider a vector $\boldsymbol{Y}=(Y_{v}, v\in V)$ comprising random variables indexed by the set $V={1, 2, \ldots, p}$, representing $p$ criteria. Each $Y_{v}$ takes values from the finite set $I_{v}$, where the number of levels $\lvert I_{v}\rvert $ can vary across different $v\in V$. When classifying $n$ observations based on these $p$ criteria, the resulting counts can be organized into a $p$-dimensional contingency table, represented by

\begin{equation*}
I=\bigtimes_{v\in V} I_{v},
\end{equation*}

\noindent where $I$ denotes the set of cells $i=(i_{v}, v\in V)$ and $i_{v}\in I_{v}$. When $Y_{v}$, $v\in V$, takes the value $i_{v}$, we will denote it as $y=i$. The number of observations for cell $i$ is denoted by $n(i)$, and the probability of an object being observed in cell $i$ is denoted by $p(i)$. If $D\subset V$, the $D$-marginal table can be written as

\begin{equation*}
I_{D}=\bigtimes_{v\in D} I_{v}.
\end{equation*}

\noindent This expression represents the collection of $D$-marginal cells $i_{D}=(i_{v},v\in D)$. For a given marginal cell $i_{D}\in I_{D}$, the count of $D$-marginal cells is denoted by $n_{D}(i_{D})=\sum_{i^{\prime}\in I; i_{D}=i^{\prime}_{D}}n(i^{\prime})$. Considering $n=\sum_{i\in I} n(i)$, we assume the cell counts $\left(n(i), i\in I \right)$ follow a multinomial distribution with the probability density function

\begin{equation}
p(n(i), i\in I)= \frac{n!}{\prod_{i\in I} n(i)!}\prod_{i\in I} p(i)^{n(i)}. 
\label{multinomial.distribution}
\end{equation}
Next we introduce the notion of hierarchical model. Here we use a subset of $V$ to denote an interaction effects involving all the vertices in the subset. For example, if the subset is $\{a,b,c\}$, we use this subset to correspond to a three-way interaction effect between vertices $a, b$ and $c$. In a hierarchical model, if a higher-order interaction is included, all the lower-order interactions will be included as well. For example, if the subset $\{a,b,c\}$ is included in the model, then all the sets $\{a\},\{b\}, \{c\}, \{a,b\}, \{b,c\},\{a,c\}$ are included in the model.  Let $\Delta$ denote a nonempty collection of subsets of $V.$ The hierarchical property dictates that if $D\in \Delta$, and $D_{1}$ is a non-empty subset of $D$, then $D_{1}\in \Delta$. We refer $\Delta$ as the generating class of the model, which essentially is a collection of subsets corresponding to a collection of different interaction terms.



We arbitrarily select an element in each $I_{v}$ to serve as the baseline level, denoted by 0. With a slight abuse of notation, we also represent by $0$ the cell in $I$ where all its levels are set to $0$. 
We define the support of a cell as
$
S(i)=\{v\in V, i_{v}\neq 0\}
$
and use $J=\{j\in I,S(j)\in \Delta\}$ to denote the subset of $I$ corresponding to the index of the generating class. 
To streamline the notation further, we introduce the symbol $j\triangleleft i$ to indicate that $S(j)\subseteq S(i)$ and $j_{S(j)}= i_{S(j)}$, indicating that \textit{$j$ is to the left of $i$}.
With this, we can model $\log p(i)$ in terms of the free parameters ${\theta_{j}, j\in J}$, yielding:

\begin{equation}
\log p(i)=\theta_{\emptyset}+\sum_{j\triangleleft i}\theta_{j},\label{log.pi}
\end{equation}

\noindent where $\theta_{\emptyset}$ is a unique value such that $\sum_{i\in I}p(i)=1$. 


To clarify the notation, let $V=\{a, b, c\}$, $\Delta=\{a, b, c, ab, bc\}$  and assume $I_{a}=I_{b}=I_{c}=\{0, 1\}$. The set $J=\{100, 010, 001, 110, 011\}$ is the set of indices associated with the free parameters. For example, if $i=101$ the subset of elements $j\in J$ such that $j\triangleleft i$ is $\{100, 001\}$. Similarly, when $i=111$ the corresponding subset is $\{100, 010, 001, 110, 011\}$, and so on. Using the unique representation (\ref{log.pi}), we express the following relationships for these two cells as
\begin{equation*}
\begin{split}
\log [p(101)/p(000)]& =\theta_{100}+\theta_{001},\\
\log [p(111)/p(000)]& = \theta_{100}+\theta_{010}+\theta_{001}+\theta_{110}+\theta_{011}.\\
\end{split}
\end{equation*}

\subsection{Covariate-dependent discrete graphical model}\label{sec:2.2}
Up to now, the hierarchical discrete graphic model is a static model which does not take into account of any covariate which may affect the cell probabilities of the contingency table. In the following subsection, we propose modelling the cell probabilities using a covariate-dependent hierarchical log-linear model, where the log-linear parameter depends on a set of covariates.

First, let us start with one covariate $x.$ We consider two graphs, the baseline graph with the generating class $J_0$ and the slope graph with the generating class $J_1$, that is,
\begin{equation}
\log( p(i)/p(0)|x)=\sum_{j\triangleleft i, j\in J_0}\theta_{j,0}+x \sum_{j\triangleleft i, j\in J_1} \theta_{j,1},\label{log.pi_example}
\end{equation} with $0$ denoting the cell with all factors at the baseline level. We can extend this to multiple covariates $x=(x^1, \dots, x^H)$, that is,
\begin{equation}
\log( p(i)/p(0)|x)=\sum_{j\triangleleft i, j\in J_0}\theta_{j,0}+\sum_{h=1}^H  x^h \sum_{j\triangleleft i, j\in J_h} \theta_{j,h}.\label{log.pi_extend}
\end{equation}


\noindent 
Let $\boldsymbol{\theta}_{0}=\Bigl(\theta_{j,0}, j\in J_0\Bigr)$ represent the vector of baseline graph free parameters, and $\boldsymbol{\theta_{h}}=\Bigl(\theta_{j,h}, j\in J_h\Bigr)$ represents the vector of slope graph parameters associated with the $h_{th}$ covariate.
Let $F_h$, $h=0,1,\cdots,H$, be the $|I|\times |J_h|$ design matrix, where the $i_{th}$ row is $\boldsymbol{f}_{h,i}^{t}$ such that $\boldsymbol{f}_{h,i}=\sum_{j\in J_h, j\triangleleft i} \boldsymbol{e}_{j}$ and $\left(\boldsymbol{e}_{j}\right)_{j\in J_h}$ is the canonical basis of $\mathbb{R}^{|J_{h}|}$. We denote the transpose of a vector or matrix using the superscript $t$.  

\begin{definition}(Covariate-dependent discrete graphical model)
Given a $p$-dimensional random vector $\boldsymbol{Y}=(Y_{v}, v\in V)$ which are indexed by a vertices set $V$, and a set of covariates $x=(x^1,x^2,\cdots,x^H)$, we define a baseline model structure with generating class $J_{0}$, and $H$ slope model structures with corresponding generating classes $J_{1},J_{2},\cdots, J_{H}$. 
Then we say $\boldsymbol{Y}$ given $x$ follows a covariate dependent discrete graphical models if 

\begin{equation}
\log( \frac{p(y=i|x)}{p(y=0|x)})=\langle \boldsymbol{\theta}_{0}, \boldsymbol{f}_{0,i}  \rangle + \sum_{h=1}^H\langle x^{h}\boldsymbol{\theta}_{h}, \boldsymbol{f}_{h,i}  \rangle = \sum_{h=0}^H\langle x^{h}\boldsymbol{\theta}_{h}, \boldsymbol{f}_{h,i}  \rangle,
\label{log.ratio}
\end{equation}
where $x^0 = 1$.
\end{definition}
We define $\boldsymbol{\theta}=(\boldsymbol{\theta}_h,h=0,1,\cdots, H)$ as the parameters in the covariate-dependent discrete graphical models.
For one observation, the likelihood function of random vector $\boldsymbol{Y}$ occurs in cell $i$ is given by 
\begin{equation}
\label{eq:likelihood_one}
L(\boldsymbol{\theta};y=i)=\exp\{\sum_{h=0}^H\langle x^{h}\boldsymbol\theta_{h}, \boldsymbol{f}_{h,i} \rangle-\log\sum\limits_{i \in I} e^{\sum_{h=0}^H\langle x^{h}\boldsymbol\theta_{h}, \boldsymbol{f}_{h,i} \rangle}\}.
\end{equation}
Given $n$ observations $\boldsymbol{Y}_{(m)}=(Y_{(m),v}, v\in V),$ $m=1,\dots,n.$ We define $\boldsymbol{t}_{h,m}=\boldsymbol{f}_{h,i}$ whenever observation $\boldsymbol Y_{(m)}$, $m=1,2,\cdots,n$, occurs in cell $i$, and $\boldsymbol{t}_{h}=\sum_{m=1}^{n} x_{m}^{h}\boldsymbol{t}_{h,m}$. For $n$ observations, the joint likelihood function is given by 
\begin{equation}
\label{eq:likelihood_n}
L_{n}(\boldsymbol{\theta})=\exp\{\sum_{h=0}^{H}\langle \boldsymbol\theta_{h}, \boldsymbol{t_h} \rangle-\sum_{m=1}^{n}\log\sum\limits_{i \in I} e^{\sum_{h=0}^{H}\langle \boldsymbol\theta_{h}, x^{h}_m\boldsymbol{f}_{h,i} \rangle}\}.
\end{equation}
The joint log-likelihood is formulated as
\begin{equation}
    \begin{split}
        \ell_{n}(\boldsymbol{\theta})&= \sum_{h=0}^{H}\langle \boldsymbol{\theta}_{h},\boldsymbol{t}_{h}\rangle-\sum_{m=1}^{n}\log \sum_{i\in I}e^{\sum_{h=0}^{H}\langle \boldsymbol{\theta}_{h},x_{m}^{h}\boldsymbol{f}_{h,i}\rangle},\\ \label{log.likelihood}
    \end{split}
\end{equation}
\noindent where (\ref{log.likelihood}) demonstrates that the covariate-dependent discrete graphical model belongs to the exponential family with $\boldsymbol{\theta}_{h}$ being the canonical parameters and $\boldsymbol{t}_{h}$ being the sufficient statistics.
As a result, parameter estimation, structure learning, and statistical inference become straightforward and manageable.

\subsection{Estimation procedure}

Let $z_{i}=\sum_{h=0}^{H}\langle \boldsymbol{\theta}_{h}x_{m}^{h},\boldsymbol{f}_{h,i}\rangle$. For $j,k\triangleleft i$, we define the following expressions:

\begin{equation}
\begin{split}
    P_{j,h}(\boldsymbol{\theta}|x_{m})&=\frac{\sum_{i\in I}\left (e^{z_{i}}\cdot f_{h,i,j}\cdot x_{m}^{h}\right)}{\sum_{i\in I}e^{z_{i}}}, \label{marginal.prob.jh}
\end{split}
\end{equation}

\noindent and

\begin{equation}
\begin{split}
    P_{j,k,h,h^{\prime}}(\boldsymbol{\theta}|x_{m})&=\frac{\sum_{i\in I}\left (e^{z_{i}}\cdot f_{h,i,j}\cdot f_{h^{\prime},i,k}\cdot x_{m}^{h}\cdot x_{m}^{h^{\prime}} \right)}{\sum_{i\in I}e^{z_{i}}}, \label{marginal.prob.jkh}
\end{split}
\end{equation}

\noindent where $h, h^{\prime}\in \{1,\cdots ,H\}$ denote two covariates.

We express the first derivative of (\ref{marginal.prob.jh}) with respect to the parameter $\theta_{k,h}$ 
as follows:

\begin{equation}
\begin{split}
&\frac{\partial P_{j,h}(\boldsymbol{\theta} |x_{m})}{\partial \theta_{k,h}} = P_{j,k,h,h^{\prime}}(\boldsymbol{\theta}|x_{m})-P_{j,h}(\boldsymbol{\theta}|x_{m})\cdot P_{k,h^{\prime}}(\boldsymbol{\theta}|x_{m}) .\\  \label{first.derivative.Pj}
\end{split}
\end{equation}
Taking the partial first derivatives of (\ref{log.likelihood}) yields the expression for the $j_{th}$ entry of the score vector:

\begin{equation}
\begin{split}
    \frac{\partial \ell_n(\boldsymbol{\theta})}{\partial \theta_{j,h}}&=\sum_{m=1}^{n}x_{m}^{h}t_{j,h,m}-\sum_{m=1}^{n} P_{j,h}(\boldsymbol{\theta}|x_{m}). \label{score.vector.h}
\end{split}
\end{equation}
Computing the partial second derivatives of (\ref{log.likelihood}) with respect to $\theta_{j,h}$ and $\theta_{k,h^{\prime}}$, gives the corresponding $(j,k)_{th}$ entry of the Hessian matrix as:
\begin{equation}
\begin{split}
\frac{\partial^{2} \ell_n(\boldsymbol{\theta})}{\partial \theta_{j,h}\partial \theta_{k,h^{\prime}}}& =-\sum_{m=1}^{n}\left[P_{j,k,h,h^{\prime}}(\boldsymbol{\theta}|x_{m})-P_{j,h}(\boldsymbol{\theta}|x_{m})\cdot P_{k,h^{\prime}}(\boldsymbol{\theta}|x_{m})\right].\label{hessian.entry.0}\\
\end{split}
\end{equation}

We can use the Newton algorithm to compute the maximum likelihood estimate based on the first derivative in \eqref{score.vector.h} and the second derivative in \eqref{hessian.entry.0} of the log-likelihood.



\subsection{Theoretical properties}

We use $\ell_{(m)}(\boldsymbol{\theta})$ to denote the log-likelihood for a single observation $\boldsymbol{Y}_{(m)}$. Let $\ell_{(m)}^{(t)}\left(\boldsymbol{\theta}\right)$ be the $t_{th}$ derivative of individual log-likelihood with respect to $\boldsymbol{\theta}$. Let $\boldsymbol{\theta}^{*}$ denote the true parameter value. Then, we write the information matrix as
$$I_n(\boldsymbol{\theta}) = \sum_{m=1}^n I_m(\boldsymbol{\theta})=\sum_{m=1}^n\operatorname{Var}_\theta[\ell_{(m)}^{(1)}\left(\boldsymbol{\theta}\right)]=\sum_{m=1}^n \left\{-\mathbb{E}\left[\ell_{(m)}^{(2)}\left(\boldsymbol{\theta}\right)\right]\right\}.$$

\begin{assumption}
Assume that  $n^{-1}I_n(\boldsymbol{\theta}^*) \xrightarrow{p}I(\boldsymbol{\theta}^*) $ as $n$ approaches infinity for some positive definite  $I(\boldsymbol{\theta}^*)$.
\end{assumption}
\begin{theorem} (Asymptotic Normality).
Under Assumptions 1, there exists a local maximizer $\hat{\boldsymbol{\theta}}$ of $\ell_n(\boldsymbol{\theta})$ such that $\left\|\hat{\boldsymbol{\theta}}-\boldsymbol{\theta}^*\right\|_2=O_p\left(n^{-\frac{1}{2}}\right)$. Furthermore,
the maximum likelihood estimator $\hat{\boldsymbol{\theta}}$ satisfies the asymptotic normality
$$
\sqrt{n}\left(\hat{\boldsymbol{\theta}}-\boldsymbol{\theta}^*\right) \xrightarrow{d} N\left(\mathbf{0}, I(\boldsymbol{\theta}^*)^{-1}\right) \text { as } n \longrightarrow \infty .
$$
\label{Normality}
\end{theorem}
Within the context of covariate-dependent discrete graphical models, we can devise various hypothesis tests on the parameters. 
For example, to evaluate whether the covariate influences the graph structure 
we consider the null hypothesis $H_0:\boldsymbol{\theta}_{h} = \boldsymbol{0}$ for any $h>0$. To access the equal impacts from the different covariates, we adopt a homogeneous
null hypothesis $H_0:\boldsymbol{\theta}_{1} = \boldsymbol{\theta}_{2} =\cdots =\boldsymbol{\theta}_{H}$.
Additionally, we can consider a null hypothesis for a single parameter that corresponds to the edge $(a,b)$ in the discrete graphical model. This null hypothesis is defined by $H_0:\theta_{j,h} = 0$, $j=(a,b)$. Such a test can be employed to evaluate the effect of the $h_{th}$ covariate's on the edge $(a,b)$ within the discrete graphical model.

\section{Pseudo-likelihood Method for High-dimensional Covariate-dependent Discrete Graphical Models}\label{sec:3}

\subsection{Methodology}
The computation of the MLE in graphical models can be an intractable problem when the dimension is high. In this subsection, we introduce a widely used MLE succedaneum: pseudo-likelihood MLE, which maximizes the pseudo log-likelihood based on conditional probability:
\[
\ell^p(\boldsymbol{\theta})=\sum_{m=1}^n\sum_{v \in V} \log p(y_{(m),v}|y_{(m),V\setminus v},x_m).
\] 
Next, we will derive the formula of the conditional probability of $y_v=i_v$ given $y_{V\setminus v}=i_{V\setminus v},$ in which we suppress the index $m$ for simplicity: 
\[
\begin{array}{lcl}
   p(y_v=i_v| y_{V\setminus v}=i_{V\setminus v},x) &=& \frac{p(y=i|x)}{p(y_{V\setminus v}=i_{V\setminus v}|x)}  =\frac{p(y=i|x)}{\sum_{i^{'}_v\in I_v}p(y_v=i^{'}_v,y_{V\setminus v}=i_{V\setminus v}|x)}\\[20pt]
     &=& \frac{\exp\{\sum_{h=0}^H\langle x^{h}\boldsymbol\theta_{h}, \boldsymbol{f}_{h,i} \rangle\}}{ \sum_{i'}\exp\{\sum_{h=0}^H\langle x^{h}\boldsymbol\theta_{h}, \boldsymbol{f}_{h,i'} \rangle \}}
\end{array}, 
\]
where $i^{'}=(i'_v,i_{V\setminus v}).$
 We break down the parameter vector $\boldsymbol\theta_h$ and $\boldsymbol{f}_{h,i}$ vectors into two parts. Here, we introduce some new notation: let $\boldsymbol\theta^v_h$ denote the subparameter vector corresponding to the variable $y_v$: $\boldsymbol\theta^v_h=(\theta_{j,h}, v\in S(j), j\in J_h)$, and let $\boldsymbol{f}_{h,i}^v$ denote the subvector of $\boldsymbol{f}_{h,i}$ corresponding to $y_v$: $\boldsymbol{f}_{h,i}^v=\sum_{j\in J_h, v\in S(j), j\triangleleft i} \boldsymbol{e}_{j}$ and $\boldsymbol{e}_{j}$ is the canonical basis vector of $\mathbb{R}^{|J^v_{h}|}$, where $J^v_h=(j\in J_h, v\in S(j))$. Then, the conditional probability becomes
\[
p(y_v=i_v| y_{V\setminus v}=i_{V\setminus v},x)=\frac{\exp\{\sum_{h=0}^H\langle x^{h}\boldsymbol\theta^{v}_{h}, \boldsymbol{f}^{v}_{h,i} \rangle \}}{ \sum_{i'}\exp\{\sum_{h=0}^H\langle x^{h}\boldsymbol\theta^{v}_{h}, \boldsymbol{f}^{v}_{h,i'} \rangle \}}.
\]
If $y_v$ takes binary values $\{0,1\}$, then we get
\[
p(y_v=0| y_{V\setminus v}=i_{V\setminus v},x)=\frac{1}{ 1+\exp\{\sum_{h=0}^H\langle x^{h}\boldsymbol\theta^{v}_{h}, \boldsymbol{f}^{v}_{h,i,i_v=1} \rangle \} }  ,
\]
and
\[
p(y_v=1| y_{V\setminus v}=i_{V\setminus v},x)=\frac{\exp\{\sum_{h=0}^H\langle x^{h}\boldsymbol\theta^{v}_{h}, \boldsymbol{f}^{v}_{h,i,i_v=1} \rangle \}}{ 1+\exp\{\sum_{h=0}^H\langle x^{h}\boldsymbol\theta^{v}_{h}, \boldsymbol{f}^{v}_{h,i,i_v=1} \rangle \} } ,
\]
then
\[
\log\frac{p(y_v=1| y_{V\setminus v}=i_{V\setminus v},x)}{p(y_v=0| y_{V\setminus v}=i_{V\setminus v},x)}=\sum_{h=0}^H\langle x^{h}\boldsymbol\theta^{v}_{h}, \boldsymbol{f}^{v}_{h,i,i_v=1}\rangle,
\]
which becomes a logistic regression problem. If $y_v$ takes more than two values, then the pseudo-likelihood MLE problem becomes a multinomial logistic regression problem.

\subsection{High-dimensional dynamic Ising model}
\label{High}

In this section, we consider a special case of the covariate-dependent discrete graphical model, 
the Ising model, in which only one-way and two-way interactions are included in the model.  
The Ising model 
is a mathematical model in statistical mechanics used to describe phase transitions in systems composed of discrete variables. These variables typically represent magnetic dipole moments of atomic spins that can be in one of two states: $1$ (up) or $0$ (down). 

In previous sections, we analyzed discrete data structured as contingency tables, where observations are categorized into cells based on
$p$ criteria or variables. Here, we transition to $n\times p$ data matrix format, where each row represents an observation, and each column corresponds to a variable. This format organizes $n$ observations of a
$p$-dimensional random vector, providing a tabular view of the dataset. The two formats---contingency tables and data matrices---are inherently interchangeable:
\begin{itemize}
\item \textbf{From data matrix to contingency table}: Each row in the
$n\times p$ matrix maps to a cell in the contingency table, determined by the combination of variable values for that observation.
\item \textbf{From contingency table to data matrix}: A cell count $n(i) (\textrm{for cell }i \in I)$ translates to $n(i)$ identical rows in the data matrix, where each row replicates the variable values $y_v=i_v$ for $v=1,\ldots , p$.
\end{itemize}
To align with the framework of Ising models, we will adjust our notations and adopt $n\times p$ data matrix format. Ising models focus on two key parameter types: main effect parameters and two-way interaction parameters. 
Let $G=(V,E)$ denote an Ising model graph structure and $\boldsymbol{Y}=(Y_v, v\in V)$ denote the random vector. Each variable $Y_v$ is indexed by vertex set $V$ and takes binary values $\{0,1\}$. The corresponding Ising model parameters are $\{\theta_v, v\in V, \theta_{uv}, (u,v)\in E\}$.
The probability mass function of Ising models takes the following format:
\begin{equation}
    \label{eq:ising}
    f(y)=\frac{1}{z(\boldsymbol{\theta})} \exp \{ \sum_{v\in V}\theta_vy_v+\sum_{(u,v)\in E}\theta_{uv}y_uy_v\},
\end{equation}
where $z(\boldsymbol{\theta})=\sum_{y\in I}\exp \{ \sum_{v\in V}\theta_vy_v+\sum_{(u,v)\in E}\theta_{uv}y_uy_v\}$ is the normalization constant.

Next, we extend the idea of covariate-dependent discrete graphical model to the Ising model. We present the detailed definition of dynamic Ising model as follows:

\begin{definition}(Dynamic Ising model)

Given a $p$-dimensional binary random vector $\boldsymbol{Y}$ which are indexed by a vertices set $V$ and a set of covariates $x=(x^1,x^2,\cdots,x^H)$, we define a baseline model structure $G_0=(V,E_0)$, and $H$ slope model structures $G_h=(V,E_h),h=1,2,\cdots H$. Then we say $\boldsymbol{Y}$ given $x$ follows a dynamic Ising model, if the probability mass function of $\boldsymbol{Y}$ given $x$ is
\begin{equation}
    \label{eq:dynamicising}
    f(y|x)=\frac{1}{z(\boldsymbol{\theta})} \exp \{\sum_{h=0}^{H}\theta_{v,h} x^hy_v+\sum_{h=0}^{H}\sum_{(u,v)\in E_h}\theta_{uv,h} x^hy_uy_v\},
\end{equation}
where $\boldsymbol{\theta}=(\theta_{v,h} , \theta_{uv,h}, v\in V, (u,v) \in E_h ,h=0,1,2,\cdots,H).$
\end{definition}
After we define a dynamic Ising model, we use the combined model structure $G=(V,\cup_{h=0}^H E_h)$ as the dynamic model structure.


\subsection{ Parameter estimation for high-dimensional dynamic Ising model}
\label{pseudo}

In this subsection, we will show how to use pseudo-likelihood method to perform parameter estimation when the high-dimensional dynamic Ising model structure is known. We will show that the dynamic Ising model parameters can be estimated in a more convenient way via logistic regression after a careful setup of design matrix.



In the dynamic Ising model structure $G$, the neighbourhoods of each vertex $v$ can be denoted as 
\[
U^h_v=\{u|(u,v)\in E_h\}, h=0,\cdots,H.
\]
We can then derive the conditional probability of $Y_v|Y_{V\setminus v}$ as follows:
\begin{equation}
p(y_v|y_{V\setminus v},x)= \frac{\exp\{\sum_{h=0}^H\theta_{v,h}x^hy_v+\sum_{h=0}^H\sum_{u\in U^h_v}\theta_{uv,h}x^hy_{u}y_v\}}{1+\exp\{\sum_{h=0}^H\theta_{v,h}x^h+\sum_{h=0}^H\sum_{u\in U^h_v}\theta_{uv,h}x^hy_{u}\}}.
\end{equation}
The overall pseudo log-likelihood function is then
\begin{equation}
        \sum_{m=1}^n \sum_{v\in V} \log p(y_{(m),v}|y_{(m),V\setminus v},x_m).
\end{equation}
Instead of maximizing the overall pseudo log-likelihood function, which is computational intensive, we adopt the approach of parallel computing. We estimate parameters from each $v_{th}$ component, and then take the average of parameter estimates to get the overall pseudo-likelihood estimates. Let $\boldsymbol{\theta}^v=(\theta_{v,h} , \theta_{uv,h},u \in U^h_v, \ h=0,1,\cdots,H)$ denote the parameters appearing in the $v_{th}$ component of the pseudo log-likelihood function, then we have:
\begin{equation}
    \begin{split}
        \ell^v(\boldsymbol{\theta}^v) &=  \sum_{h=0}^H(\theta_{v,h}\sum_{m=1}^nx^h_{m}y_{(m),v}+\sum_{u \in U^h_v}\theta_{uv,h} \sum_{m=1}^n x^h_{m}y_{(m),u}y_{(m),v} ) \\
        &-\sum_{m=1}^n \log [1+\exp\{\sum_{h=0}^H(\theta_{v,h}x^h_{m}+\sum_{u\in U^h_v}\theta_{uv,h}x^h_{m}y_{(m),u})\}]. \\
    \end{split}
\end{equation}
Now, let 
$t_{v,h}=\sum_{m=1}^nx^h_{m}y_{(m),v},\ t_{uv,h}=\sum_{m=1}^n x^h_{m}y_{(m),u}y_{(m),v}$, we get
\begin{equation}
\label{eq:conditional1}
        \ell^v(\boldsymbol{\theta}^{v}) =  \sum_{h=0}^H(\theta_{v,h}t_{v,h}+\sum_{u \in U^h_v}\theta_{uv,h} t_{uv,h} ) -\sum_{m=1}^n \log (1+z_m(\boldsymbol{\theta}^{v})) ,
\end{equation}
where $z_m(\boldsymbol{\theta}^{v})=\exp\{\sum_{h=0}^H (\theta_{v,h}x^h_{m}+\sum_{u\in U^h_v}\theta_{uv,h}x^{h}_{m}y_{(m),u}\}$. The score functions are then
\begin{equation}
    \begin{split}
             \frac{\partial \ell^v(\boldsymbol{\theta}^{v})}{\partial \theta_{h,v}} &= t_{v,h}-\sum_{m=1}^n \frac{x^h_{m}z_m(\boldsymbol{\theta}^{v})}{1+z_m(\boldsymbol{\theta}^{v})} ,\\
                  \frac{\partial \ell^v(\boldsymbol{\theta}^{v})}{\partial \theta_{uv,h}} &= t_{uv,h}-\sum_{m=1}^n \frac{x^h_{m}y_{(m),u}z_m(\boldsymbol{\theta}^{v})}{1+z_m(\boldsymbol{\theta}^{v})}. \\
    \end{split}
    \label{eq:score}
\end{equation}

The equations of score functions can be rewritten into vector format, after we introduce a design matrix notation here. Let $\boldsymbol{t}=(t_{v,h},t_{uv,h}, u\in U^h_v, h=0,1,\cdots,H)^t$ be the vector of sufficient statistics in $\ell ^v(\boldsymbol{\theta}^v)$ and define each row of the design matrix as follows:
\begin{equation}
    \label{eq: designmatrix}
       \boldsymbol{d}_m=( x_m^h, x_m^hy_{(m),u}, u\in U^h_v, h=0,\cdots,H).
\end{equation}
 Then the design matrix $\boldsymbol{D}=(\boldsymbol{d}^t_1,\boldsymbol{d}^t_2,\cdots,\boldsymbol{d}^t_n)^t$. Let vector $\boldsymbol{P}=(\frac{z_m(\boldsymbol{\theta}^{v})}{1+z_m(\boldsymbol{\theta}^{v})}, m=1,2,\cdots, n)$ denote the conditional probability vector of the samples.  Now we can rewrite the score function in the vector format:
\begin{equation}
    \label{eq:score2}
    \frac{\partial {\ell^v(\boldsymbol{\theta}^{v})}}{ \partial \boldsymbol{\theta}^{v}}=\boldsymbol{t}-\boldsymbol{ D }^{t}\boldsymbol{ P}.
\end{equation}
Furthermore, we can derive the Hessian matrix as follows:
\begin{equation}
    H_v= -\boldsymbol{D}^{t}\text{diag}(\boldsymbol{P}) \boldsymbol{D}.
\end{equation}
With the equations of score function and Hessian matrix, one can perform the parameter estimation via Newton method. With the help of statistics software R, or other scientific software, one can also get the parameter estimation via logistic regressions using design matrix $\boldsymbol{D}$ and response vector $(y_{(m),v}, m=1,2, \cdots, n)^t$.

\section{Bayesian Model Selection for Covariate-Dependent Discrete Graphical Models}\label{sec:4}

In this section, we describe the birth-death MCMC method and present the implementation of the scalable birth-death MCMC algorithm for structure learning in high-dimensional covariate-dependent discrete graphic models. For notational convenience, we will present the algorithm in the setting of dynamic Ising model.

\subsection{Birth-death MCMC method for dynamic Ising model}
 We propose to employ a Bayesian model selection approach to learn the neighbourhood structure of each variable, and then combine the neighbourhoods of all the variables to achieve a global dynamic Ising model. We define a neighbourhood set of each variable $v$ as $N_v=\{U^h_v, h=0,1,\cdots, H\}.$ The space of this set is denoted as $\mathcal{N}_{v}.$ We assign a prior distribution $\pi(N_{v})$ over this space. Given the observed data $(x,y)$, we can compute the posterior distribution of the neighbourhood model of each variable $v$ based on the conditional likelihood as follows:
\begin{equation}
 \label{eq:post}
 p(N_{v}|y,x) \propto p(y_v|y_{V\setminus v},x)\pi(N_{v}) = p(y_v|y_{N_v},x)\pi(N_{v}),
 \end{equation}
where $y_{N_v}=(y_u, u\in N_v)$. Given the neighbourhood structure $N_v$, we assume $p(\boldsymbol{\theta^v}|N_v)$ is the parameter prior distribution associate with $N_v$, and then
\begin{equation}
p(y_v|y_{N_v},x)=\int p(y_v|y_{N_v} , x,\boldsymbol{\theta}^{v})p(\boldsymbol{\theta}^{v}|N_v)d\boldsymbol{\theta^v}.
\end{equation}
There does not exist a closed form for this posterior distribution \eqref{eq:post}. We adopt a birth-death MCMC (BDMCMC) method to simulate this posterior distribution and sample neighbourhoods from the space $\mathcal{N}_v$.

The BDMCMC is a continuous-time Markov process, which explores the space by adding and removing variables corresponding to birth and death jumps. For each $v,$ given the current neighbourhood $N_v$, the birth and death events are defined by the following independent Poisson processes:
\begin{itemize}
	\item \textbf{Birth event:} Each variable $u \not \in N_v$ is born independently of the other edges as a Poisson process with rate $B_{u}(N_v)$. If this birth event of $u$ happens, the process jumps to the new model: $N_v\cup u$.
	\item \textbf{Death event:} Each variable $u' \in N_v$ dies independently of the other edges as a Poisson process with rate $D_{u'}(N_v)$. If this death event of $u'$ happens, the process jumps to the new model: $N_v\setminus u'$.
\end{itemize}
The time to the next birth-death jump follows the exponential distribution with mean $$\lambda^{v}=\frac{1}{\sum_{u\not \in N_v}B_{u}(N_v)+\sum_{u' \in N_v}D_{u'}(N_v)},$$
and the probability of the birth and death events are respectively
\[
\begin{array}{lcl}
p_{N_v}(u) &=& \frac{B_{u}(N_v)}{\sum_{u\not \in N_v}B_{u}(N_v)+\sum_{u' \in N_v}D_{u'}(N_v)}, \quad u\not \in N_v, \\[15pt]
q_{N_v}(u') &=& \frac{D_{u'}(N_v)}{\sum_{u\not \in N_v}B_{u}(N_v)+\sum_{u' \in N_v}D_{u'}(N_v)}, \quad u' \in N_v.
\end{array}
\]
Here, we use the notation $p_{N_v}(u)$ to denote the probability of adding $u$ to neighbourhood $N_v$, that is, $p_{N_v}(u)$ is the probability of this Markov process jumping from $N_v$ to $N_v\cup u$. Similarly, $q_{N_v}(u')$ is the probability of this Markov process jumping from $N_v$ to $N_v\setminus u'$. To ensure that the BDMCMC converges to the posterior distribution \eqref{eq:post}, we give the following theorem.

\begin{theorem}
Given the BDMCMC process as described above, for any model $N_v\in \mathcal{N}_v$, and any variable $u\not \in N_v$, we define $N'_v=N_v \cup u$. If the birth/death rates are defined as follows:
\begin{equation}
\begin{array}{lcl}
B_{u}(N_v,N'_v) &=&\min ( \cfrac{p(N'_v|y,x)}{p(N_v|y,x)} ,1), \\[20pt]

D_{u}(N'_v,N_v) &=&  \min (\cfrac{p(N_v|y,x)}{p(N'_v|y,x)},1),
\end{array}
\label{eq:bdrate}
\end{equation}
 then the stationary distribution of the above BDMCMC is $\{p(N_v|y,x),N_v\in \mathcal{N}_v \}$, which is defined in equation \eqref{eq:post}.
\end{theorem}
\subsection{Birth-death rate computation}
\label{sec:bdrate}

In this subsection, we offer a fast and accurate approximation of $\cfrac{p(N'_v|y,x)}{p(N_v|y,x)}$:
\[
\cfrac{p(N'_v|y,x)}{p(N_v|y,x)}=\frac{p(y_v|y_{N'_v},x)}{p(y_v|y_{N_v},x)}\times \frac{\pi(N'_v)}{\pi(N_v)}.
\]
The probability $p(y_v|y_{N_v},x)$ can be estimated by the Bayesian information criterion (BIC) or the extended BIC:

\[
BIC(N_v)=-2\ell^v(\hat{\boldsymbol{\theta}}^v)+p_v\log(n),
\]
or
\[
EBIC(N_v)=-2\ell^v(\hat{\boldsymbol{\theta}}^v)+p_v\log(n)+2\omega p_v \log(p), \omega\geq 0,
\]
respectively, where $\hat{\boldsymbol{\theta}}^v=\arg \max \ell^v(\boldsymbol{\theta}^v)$, $p_v$ is cardinality of $N_v$, $p$ is the dimension of $y$, and $\omega$ is the tunning parameter. Then
\[
p(y_v|y_{N_v},x)\approx \exp \{-BIC(N_v)/2\},
\]
or
\[
p(y_v|y_{N_v},x)\approx \exp \{-EBIC(N_v)/2\}.
\]
The computation of the birth-death rates now is reduced to the computation of $\hat{\boldsymbol{\theta}}^v$  and pseudo log-likelihood function value $\ell^v(\hat{\boldsymbol{\theta}}^v)$.

\subsection{Scalable birth-death MCMC(SBDMCM) for high-dimensional dynamic Ising models}
\label{sec:sbdmcmc}

The previous subsection discussed how to use BDMCMC to sample neighbourhood models. For each variable $v$, the problem turns to a variable selection problem in regression, which means the method can be scaled to high-dimensional situations. After obtaining the neighbourhood models, we need to combine them together to obtain a global dynamic Ising model. Now, we refer to our method as scalable birth-death MCMC (SBDMCMC):
\begin{enumerate}

\item  Apply BDMCMC algorithm with stationary distribution $p(N_v|y,x)$ and generate
 neighbourhood model samples $\{N_{v,1},N_{v,2},N_{v,3},\cdots\}$. Use Bayesian model averaging to decide which variables are in the neighbourhood:

\[
p(u\in N_v)=\sum_{s} \mathbf{1}_{N_{v,s}}(u)p(N_{v,s}|y,x), \quad u\in V\setminus v.
\]
We set the probability value threshold as $0.5$, that is,

\[
\begin{cases}u\in N_v \quad \mbox{if}\quad p(u\in N_v)\geq 0.5, \\ u\not \in N_v \quad \mbox{if}\quad p(u\in N_v)< 0.5. \end{cases}
\]
\item  Obtain the global graph structure based on the AND or OR rule from all the neighbourhoods $N_v, v\in V$:
\begin{itemize}
\item AND rule:
\[
(u,v) \begin{cases} \in E \quad \text{if } u\in N_v \text{ and } v \in N_u, \\ \not \in E, \quad \text{otherwise}. \end{cases}
\]

\item OR rule:
\[
(u,v) \begin{cases} \in E \quad \text{if } u\in N_v \text{ or } v \in N_u, \\ \not \in E, \quad \text{otherwise}. \end{cases}
\]
\end{itemize}

\end{enumerate}

\section{Simulations}\label{sec:5}
In this section, we present a series of numerical experiments to demonstrate the effectiveness and validity of the proposed methods. These methods include MLE using the Newton method, the likelihood ratio test, and pseudo-likelihood estimation for high-dimensional settings. The results of these simulations are conducted on 100 data sets and are summarized below.

\subsection{Maximum likelihood estimation}

In the following, we present simulation studies of the maximum likelihood estimation for covariate-dependent discrete graphical models where the baseline and slope have the same structure, and when they have different structures.

\subsubsection{Baseline and slope with the same structure}\label{section: MLE}
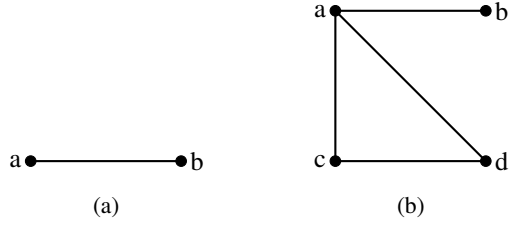
\begin{figure}[ht]
     \centering
     \begin{subfigure}[b]{0.24\textwidth}
         \centering
         \begin{tikzpicture}
\draw[black, thick] (-3, 0.25)  -- (-1, 0.25);
\filldraw[black] (-3, 0.25) circle (2pt) node[anchor=east]{a};
\filldraw[black] (-1, 0.25) circle (2pt) node[anchor=west]{b};
\end{tikzpicture}
         \caption{ }
         \label{fig:G_t}
     \end{subfigure}
     \begin{subfigure}[b]{0.24\textwidth}
         \centering
         \begin{tikzpicture}
\draw[black, thick] (-3, 2.25) -- (-3, 0.25);
\draw[black, thick] (-3, 0.25)  -- (-1, 0.25);
\draw[black, thick] (-1, 2.25)  -- (-3, 2.25);
\draw[black, thick] (-3, 2.25)  -- (-1, 0.25);
\filldraw[black] (-3, 2.25) circle (2pt) node[anchor=east]{a};
\filldraw[black] (-1, 2.25) circle (2pt) node[anchor=west]{b};
\filldraw[black] (-3, 0.25) circle (2pt) node[anchor=east]{c};
\filldraw[black] (-1, 0.25) circle (2pt) node[anchor=west]{d};
\end{tikzpicture}
         \caption{ }
         \label{fig:G_m1}
     \end{subfigure}
        \caption{Graphs (a) and (b) are visualizations of $G(2)$ and $G(4)$, respectively.}
        \label{fig:four graphs}
\end{figure}

We simulate binary data generated from two different graphs, as depicted in Figure \ref{fig:four graphs}. Graph (a), denoted as $G(2)$, is 2-dimensional and includes one two-way interaction term. Graph (b), denoted as $G(4)$, is 4-dimensional, and includes four two-way interaction terms and one three-way interaction term. In this simulation, the baseline and slope share the same graph structure. In these graphs, the cell probabilities depend on one covariate $x^{1} $ where $x_{m}^{1}$, $m=1,2,\cdots,n$, can belong to either set $S_{1} = \{0.1, 0.2, 0.3, 0.4, 0.5\}$ or set $S_{2} =  \{0.05, 0.1$, $0.15, \cdots, $0.85, 0.9, $0.95, 0.99\}$ with an increment of 0.05. For each sample size $n\in \{5000, 10000\}$, there are $n/5$ or $n/20$ observations corresponding to each number in sets $S_{1}$ and $S_{2}$, respectively.

We employ the Newton-Raphson method to estimate the vector of $\boldsymbol{\theta}$. This method leverages the score vector and its first derivative, the Hessian matrix. The estimator is denoted as $\boldsymbol{\hat{\theta}}$.
We randomly select the true parameter $\boldsymbol{\theta}^{*}$ from a normal distribution $\mathcal{N}(\boldsymbol{0},\boldsymbol{1})$. In order to evaluate the accuracy of the parameter estimation, we compute the $\ell_{2}$-norm of the error vector $\epsilon= \boldsymbol{\theta}^{*}-\boldsymbol{\hat{\theta}}$. In Table \ref{table:MLE}, we present the average $||\epsilon||_{2}$ divided by the number of parameters over 100 replications, along with the corresponding standard errors.

\begin{table*}[ht]
\caption{Estimation accuracy for two different graphs with varying sample size and covariate dimension. Standard errors are indicated in parentheses.}
\centering
\makebox[\textwidth][c]{
\begin{tabular}{ccccccccc}
\hline
\hline
 \noalign{\smallskip}
$n$    &\multicolumn{2}{c}{5000}&\multicolumn{2}{c}{10000}\\
\cmidrule(r){2-3}  \cmidrule(r){4-5} \cmidrule(r){6-7}  \cmidrule(r){8-9}
 &$x_{m}^{1}\in S_{1}$  &$x_{m}^{1}\in S_{2}$ &$x_{m}^{1}\in S_{1}$ &$x_{m}^{1}\in S_{2}$\\
  $G(2)$ &0.0867(0.0047)  &0.0461(0.0020) &0.0560(0.0019) &0.0322(0.0009)\\
\hline
$G(4)$ &0.0729(0.0066)  &0.0402(0.0028) &0.0511(0.0023) &0.0276(0.0018)\\
\hline
\hline
\end{tabular}
}
\label{table:MLE}
\end{table*}

\subsubsection{Baseline and slope with different structures}

\begin{figure}[ht]
     \centering
     \begin{subfigure}[b]{0.24\textwidth}
         \centering
         \begin{tikzpicture}
\draw[black, thick] (-3, 0.25)  -- (-1, 0.25);
\filldraw[black] (-3, 0.25) circle (2pt) node[anchor=east]{a};
\filldraw[black] (-1, 0.25) circle (2pt) node[anchor=west]{b};
\end{tikzpicture}
         \caption{ }
         \label{fig:G_20}
     \end{subfigure}
     \begin{subfigure}[b]{0.24\textwidth}
         \centering
         \begin{tikzpicture}
\filldraw[black] (-3, 0.25) circle (2pt) node[anchor=east]{a};
\filldraw[black] (-1, 0.25) circle (2pt) node[anchor=west]{b};
\end{tikzpicture}
         \caption{ }
         \label{fig:G_21}
     \end{subfigure}
        \caption{Graphs (a) and (b) are visualizations of $G(2)_{0}$ and $G(2)_{1}$, respectively.}
        \label{fig:two graphs1}
\end{figure}
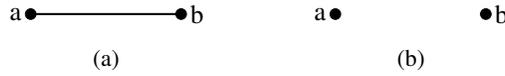

We simulate binary data generated from four distinct graphs to investigate the maximum likelihood estimation under varying baseline and slope structures. Figures \ref{fig:two graphs1} (a) and (b) depict the graphs $G(2)_{0}$ and $G(2)_{1}$, corresponding to the baseline and slope graphs, respectively. In the first model, the graphs are 2-dimensional and $G(2)_{0}$ includes one two-way interaction term.

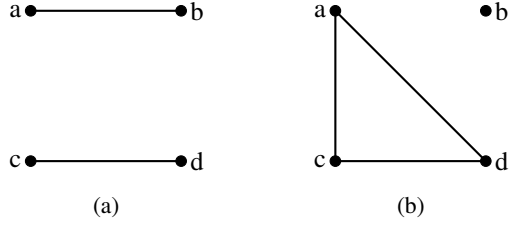
\begin{figure}[ht]
     \centering
     \begin{subfigure}[b]{0.24\textwidth}
         \centering
         \begin{tikzpicture}
\draw[black, thick] (-3, 2.25) -- (-1, 2.25);
\draw[black, thick] (-3, 0.25)  -- (-1, 0.25);
\filldraw[black] (-3, 2.25) circle (2pt) node[anchor=east]{a};
\filldraw[black] (-1, 2.25) circle (2pt) node[anchor=west]{b};
\filldraw[black] (-3, 0.25) circle (2pt) node[anchor=east]{c};
\filldraw[black] (-1, 0.25) circle (2pt) node[anchor=west]{d};
\end{tikzpicture}
         \caption{ }
         \label{fig:G_40}
     \end{subfigure}
     \begin{subfigure}[b]{0.24\textwidth}
         \centering
         \begin{tikzpicture}
\draw[black, thick] (-3, 2.25) -- (-3, 0.25);
\draw[black, thick] (-3, 0.25)  -- (-1, 0.25);
\draw[black, thick] (-3, 2.25)  -- (-1, 0.25);
\filldraw[black] (-3, 2.25) circle (2pt) node[anchor=east]{a};
\filldraw[black] (-1, 2.25) circle (2pt) node[anchor=west]{b};
\filldraw[black] (-3, 0.25) circle (2pt) node[anchor=east]{c};
\filldraw[black] (-1, 0.25) circle (2pt) node[anchor=west]{d};
\end{tikzpicture}
         \caption{ }
         \label{fig:G_41}
     \end{subfigure}
        \caption{Graphs (a) and (b) are visualizations of $G(4)_{0}$ and $G(4)_{1}$, respectively.}
        \label{fig:two graphs2}
\end{figure}

Figures \ref{fig:two graphs2} (a) and (b) illustrate the graphs $G(4)_{0}$ and $G(4)_{1}$, representing the baseline and slope graphs for a 4-dimensional model. The graph $G(4)_{0}$ contains two two-way interaction terms, while $G(4)_{1}$ includes one three-way interaction term. 

As in subsection \ref{section: MLE}, we apply the Newton-Raphson method to estimate the vector of $\boldsymbol{\theta}$, where the sets $S_{1}$ and $S_{2}$, as well as the error vector $\epsilon$ are defined. In Table \ref{table:MLE2}, we present the average $||\epsilon||_{2}$ divided by the number of parameters over 100 replications, along with the corresponding standard errors. 

\begin{table*}[ht]
\caption{Estimation accuracy for two different models with varying sample size and covariate dependence structures. Standard errors are indicated in parentheses.}
\centering
\makebox[\textwidth][c]{
\begin{tabular}{ccccccccc}
\hline
\hline
 \noalign{\smallskip}
$n$    &\multicolumn{2}{c}{5000}&\multicolumn{2}{c}{10000}\\
\cmidrule(r){2-3}  \cmidrule(r){4-5} \cmidrule(r){6-7}  \cmidrule(r){8-9}
 &$x_{m}^{1}\in S_{1}$  &$x_{m}^{1}\in S_{2}$ &$x_{m}^{1}\in S_{1}$ &$x_{m}^{1}\in S_{2}$\\
  $G(2)_{0}, G(2)_{1}$ &0.0554(0.0018)  &0.0403(0.0010) &0.0302(0.0009) &0.0224(0.0005)\\
\hline
$G(4)_{0}, G(4)_{1}$ &0.0554(0.0039)  &0.0399(0.0020) &0.0332(0.0025) &0.0235(0.0011)\\
\hline
\hline
\end{tabular}
}
\label{table:MLE2}
\end{table*}

\subsection{Likelihood ratio test}
\label{section: like}
Next, we conduct simulations to evaluate the performance of the likelihood ratio test for the null hypothesis $H_0:\boldsymbol{\theta}_{h} = \boldsymbol{0}$ for some $h>0$.
As demonstrated by the proof of Theorem \ref{Normality}, the test statistic, under the null hypothesis,
$$
-2 \log \Lambda =-2 \ell_n(\boldsymbol{\hat{\theta}}_{-h}) +2 \ell_n(\boldsymbol{\boldsymbol{\hat{\theta}}}) \rightarrow \chi_k^2
$$
in distribution as $n \rightarrow \infty$, where $k$ is the dimension of $\boldsymbol{\theta}_{h},$  $\boldsymbol{\hat{\theta}}$ is the unconstrained MLE and 
$$\boldsymbol{\hat{\theta}}_{-h} =\arg \max \limits_{\boldsymbol{\theta}_{h} = \boldsymbol{0}} \ell_n(\boldsymbol{\theta}).$$
For graphs (a) and (b) in Figure \ref{fig:four graphs}, we set covariate $x_m^{1} \in S_{2}$, which is defined in subsection \ref{section: MLE}. We construct the following hypothesis:
$$H_0:\boldsymbol{\theta}_{1} = \boldsymbol{0}\ \ \ \ \ \  \text{vs.}  \ \ \ \ \ \ H_a:\boldsymbol{\theta}_{1} \neq \boldsymbol{0}.$$ The degrees of freedom in the chi-square distribution are 3 and 9 corresponding to the graphs (a) and (b) in Figure \ref{fig:four graphs}, respectively.
To assess the powers, we consider a specific alternative situation where the slope graph parameters all take the value of $\gamma.$ 
The Type I errors and powers associated with the likelihood ratio test are presented in Table \ref{table:testing_power}.
\begin{table*}[ht]
\caption{Type I errors and Powers for different types of graphs with the significance level equal to 0.05. Standard errors are indicated in parentheses.}
\centering
\makebox[\textwidth][c]{
\begin{tabular}{ccccccc}
\hline
\hline
   &\multicolumn{2}{c}{$\gamma = 0$}&\multicolumn{2}{c}{$\gamma = 0.1$}&\multicolumn{2}{c}{$\gamma = 0.5$}\\
     \cmidrule(r){2-3}  \cmidrule(r){4-5} \cmidrule(r){6-7}
 \noalign{\smallskip}
$n$ &5000 &10000& 5000 &10000 &5000& 10000
\\
\hline
$G(2)$ &0.053(0.0071)  &0.048(0.0067) &0.489(0.0158) &0.818(0.0122)&1(0.0000)&1(0.0000)\\
\hline
$G(4)$ &0.061(0.0076)  &0.057(0.0073) &0.215(0.0130) &0.434(0.0157)&1(0.0000)&1(0.0000)\\
\hline
\hline
\end{tabular}
}
\label{table:testing_power}
\end{table*}

\subsection{Pseudo likelihood estimation}

\begin{figure}[ht]
\subfloat[Baseline graph structure $G_0$ ]
{\includegraphics[width=0.45 \linewidth]{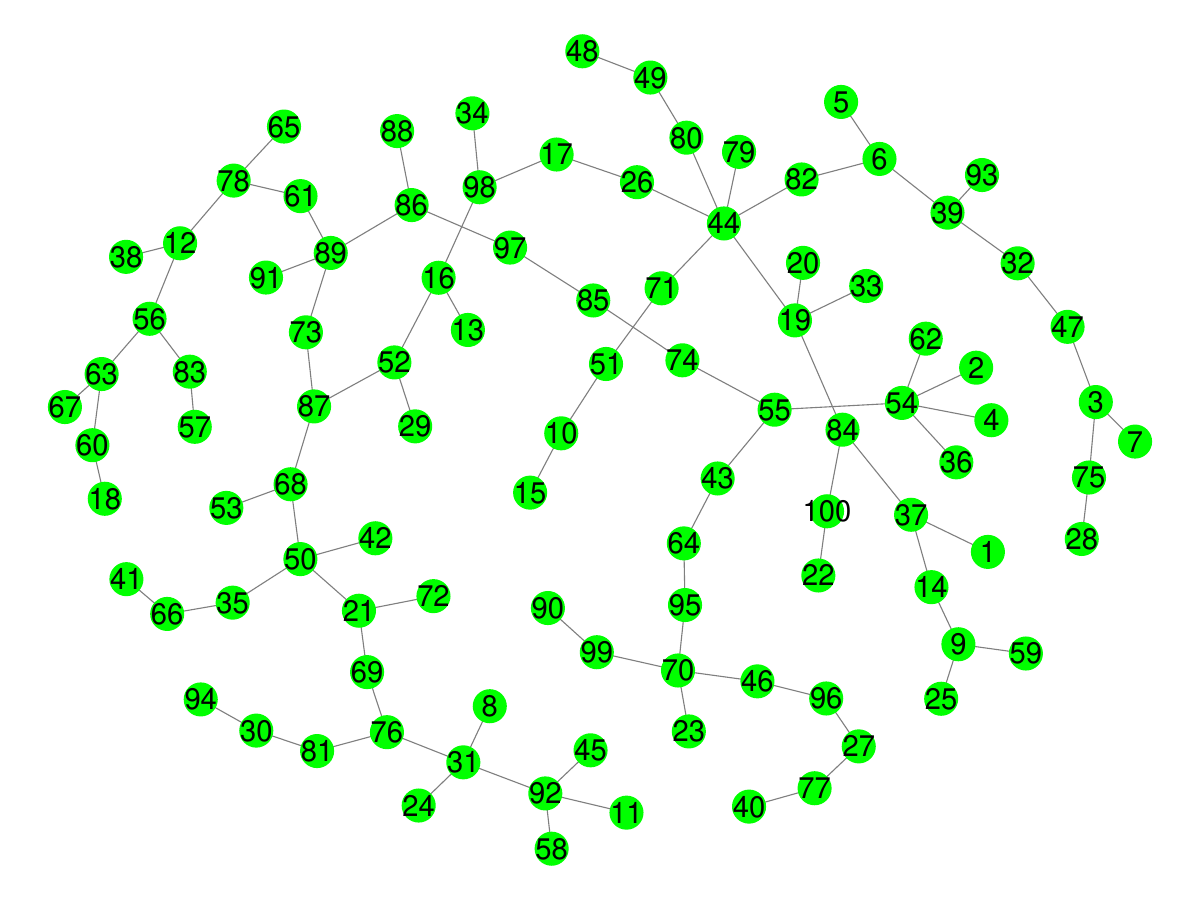} \label{fig:g0}}
\subfloat[Slope graph structure $G_1$  ]
{\includegraphics[width=0.45 \linewidth]{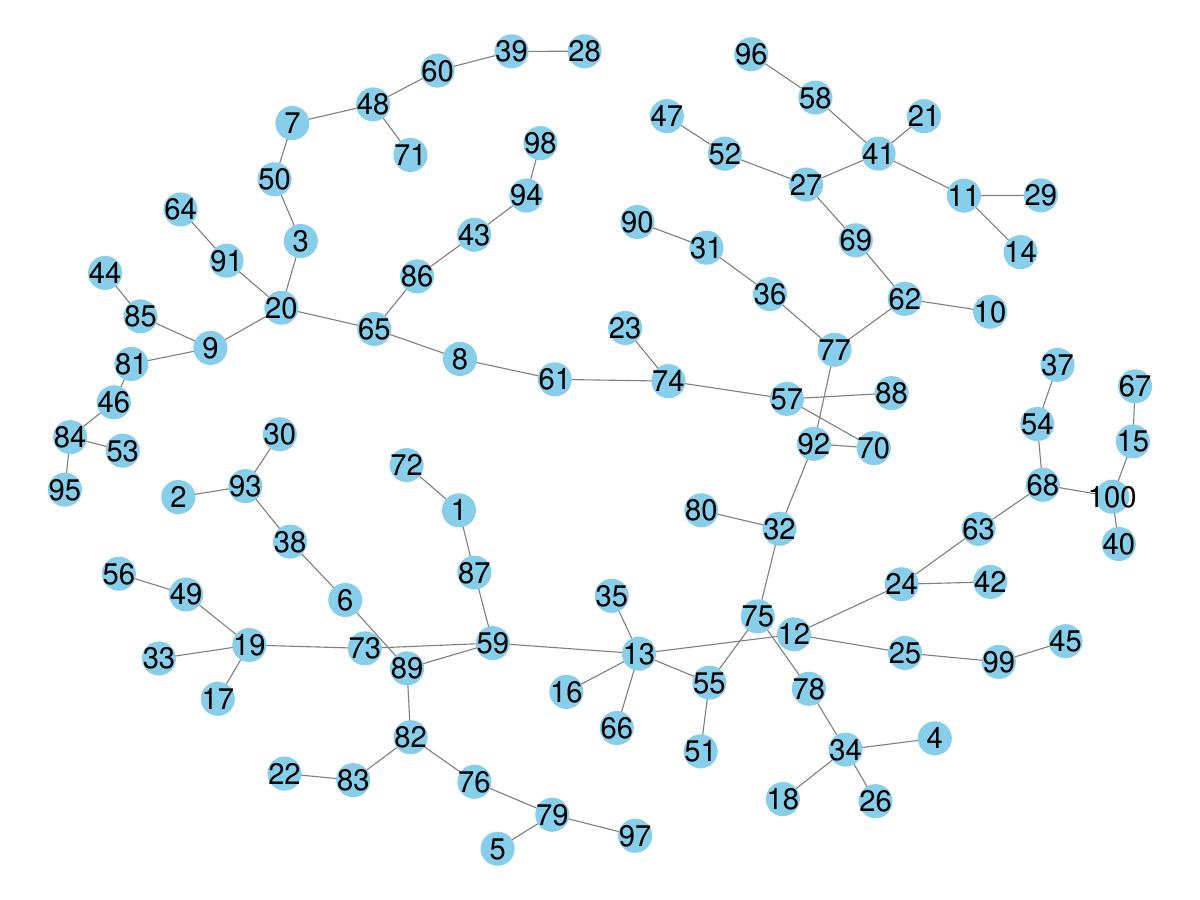} \label{fig:gh}}
\caption{Dynamic Ising model structures for the simulation study}
\label{fig:isingstruc}
\end{figure}

In this subsection, we create a dynamic Ising model with 100 variables and one covariate. The structure of the model is illustrated in Figure \ref{fig:isingstruc}.

\subsubsection{Dynamic Ising model parameter estimation via pseudo-likelihoods}

\begin{figure}[ht]
    \centering
    \includegraphics[width=0.5\linewidth]{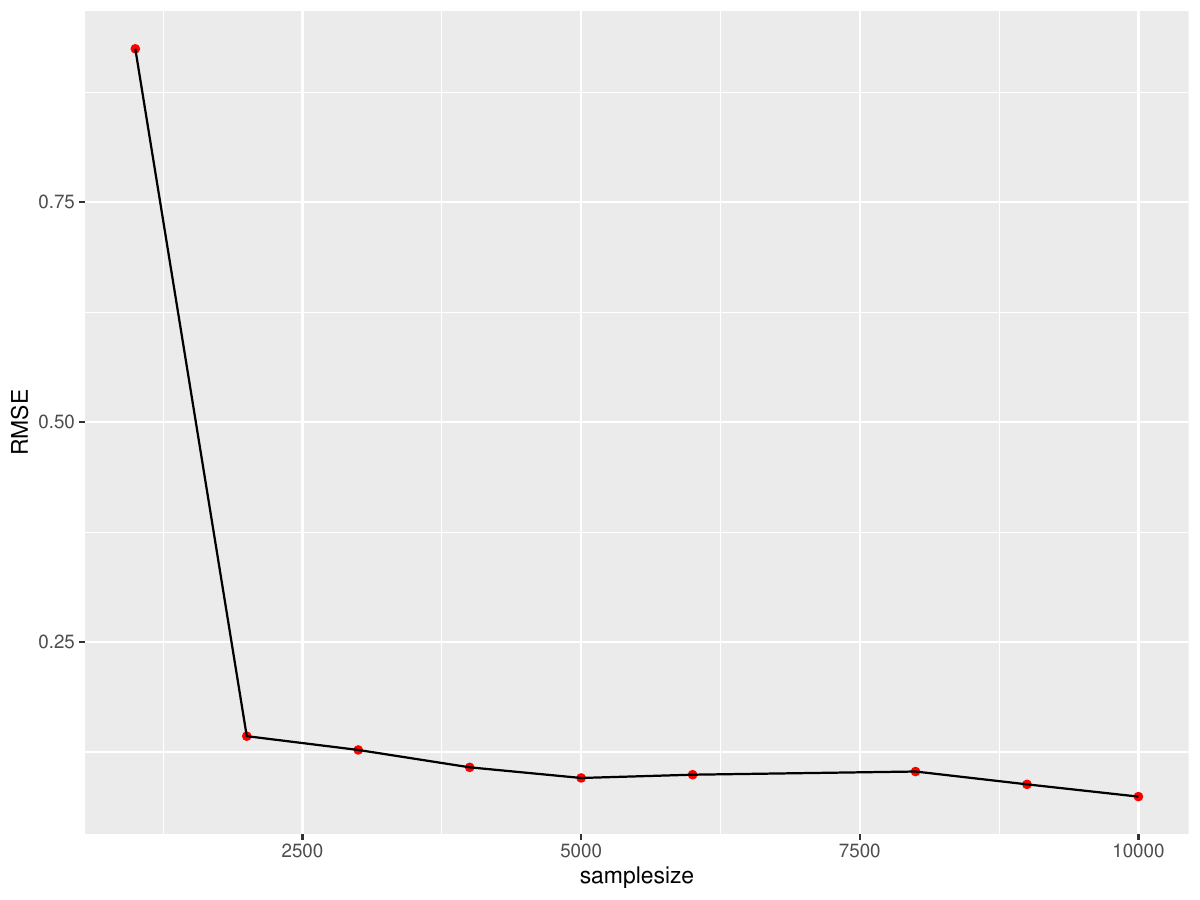}
    \caption{ Relative MSE v.s. sample size for the model in Figure \ref{fig:isingstruc}}
    \label{fig:enter-label}
\end{figure}

We generate the values of the parameters randomly, drawing from a standard normal distribution, and create 1,000,000 sample points using the Gibbs sampling method. Subsequently, we estimate the parameter values using sample sizes ranging from 10,000 to 100,000. We then calculate the relative Mean Square Error (MSE) of the parameter estimation:
\[
RMSE=\frac{\lVert \hat{\boldsymbol{\theta}}^v-\boldsymbol{\theta^{*v}}\rVert^2}{\lVert\boldsymbol{\theta^{*v}}\rVert^2}
\] and plot these values against the sample size, as depicted in Figure \ref{fig:enter-label}. The plot illustrates that the RMSE decreases as the sample size increases.

\subsubsection{Dynamic Ising model selection via SBDMCMC}
Now, we use the same simulated data from Section 5.3.1 to perform dynamic Ising model selection via SBDMCMC and compare our method with Lasso method, where the Lasso penalty parameter is selected based on BIC score. We assume the model structures are unknown, and try to learn both the base structure and slope structure.

To compare graphical model selection performance for different methods, we use the $F_1$-score, which is defined as follows:
\[
F_1-\textrm{score}=\frac{2TP}{2TP+FP+FN},
\]
where $TP$ is the true positive edges, $FP$ is the false positive edges, and $FN$ is the false negative edges. A higher $F_1$-score indicates better performance of the graphical model learning algorithm.

\begin{figure}[ht]
\subfloat[Baseline graph structure $G_0$ ]
{\includegraphics[width=0.45 \linewidth]{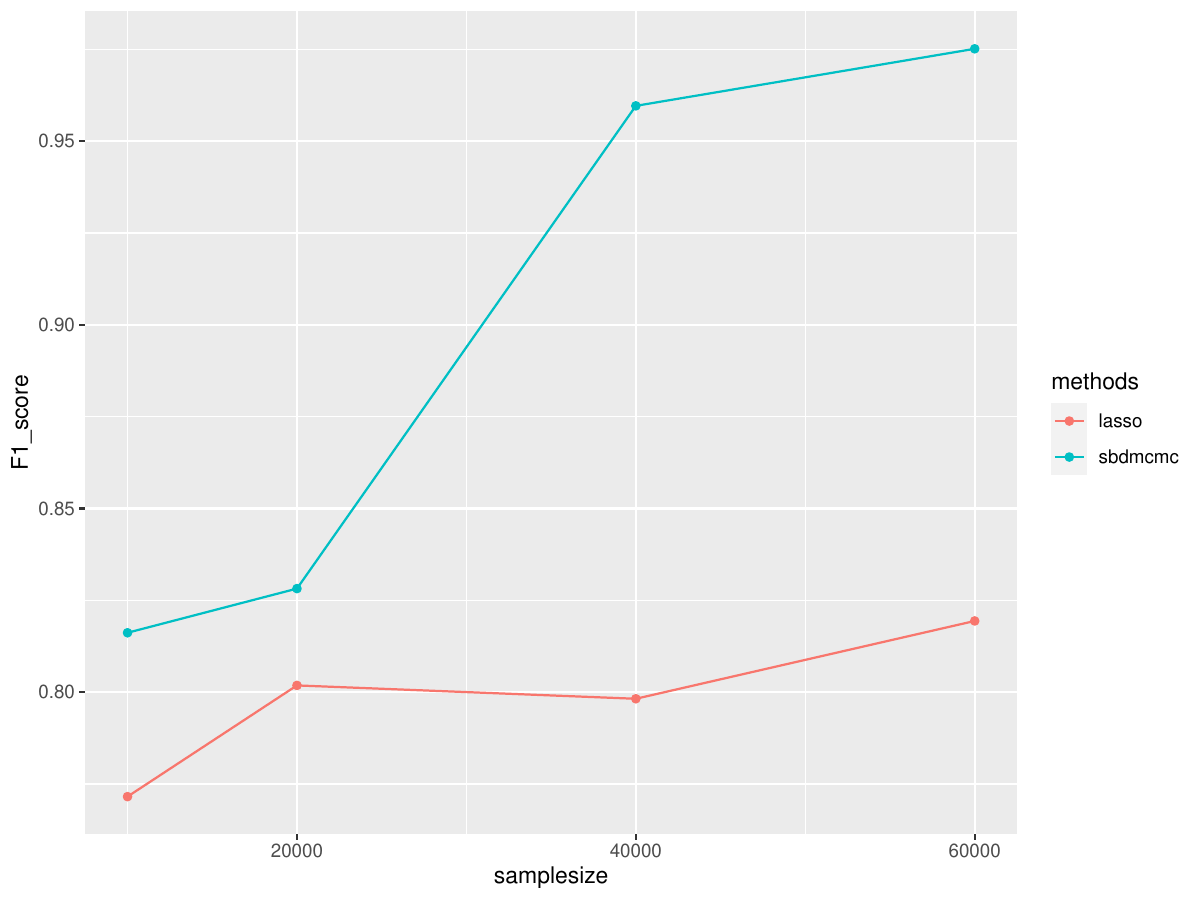} \label{fig:bf1}}
\subfloat[Slope graph structure $G_1$  ]
{\includegraphics[width=0.45 \linewidth]{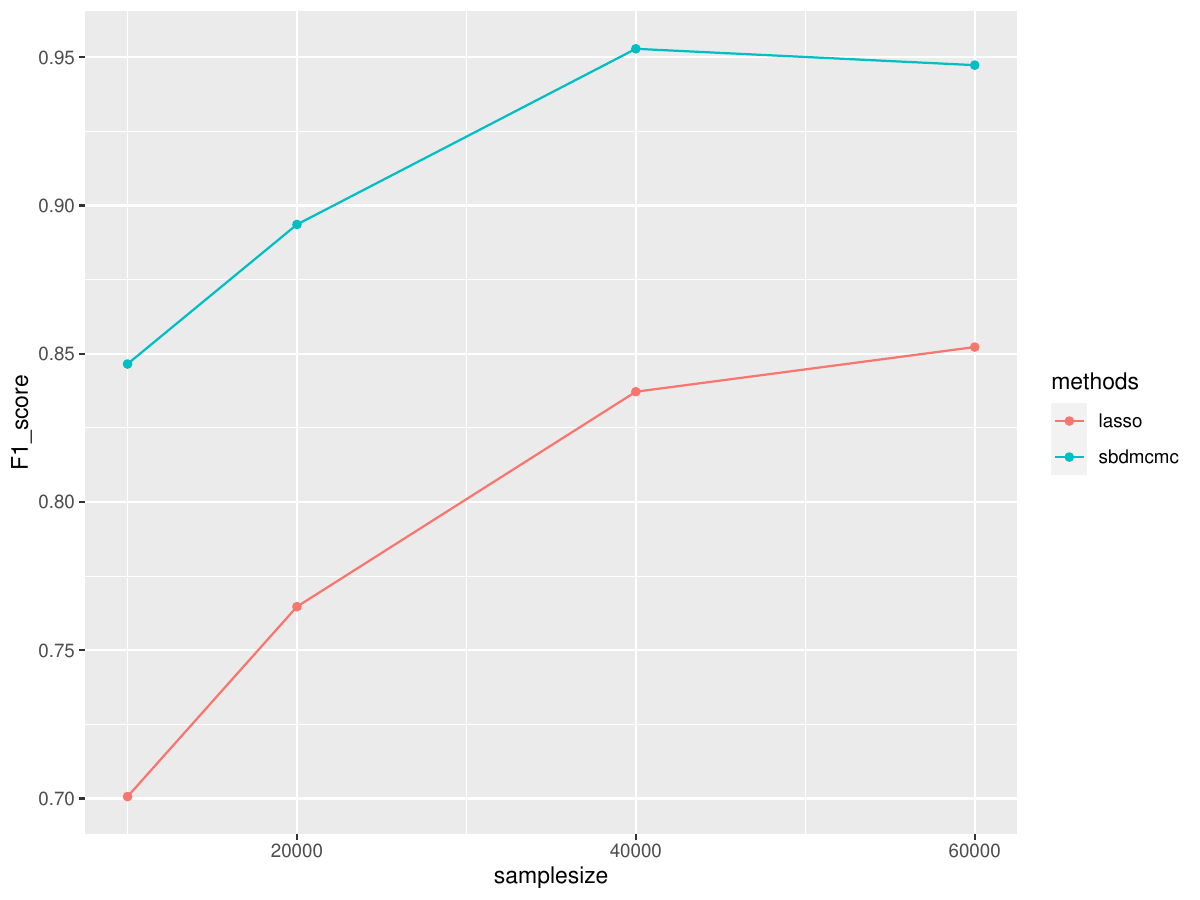} \label{fig:sf1}}
\caption{$F_1$-\textrm{score} comparison for SBDMCMC and Lasso methods}
\label{fig:f1score}
\end{figure}

Various sample sizes, as illustrated in Figure \ref{fig:f1score}, are set up to learn the dynamic Ising model structure given the dataset from Section 5.3.1. Then the $F_1$-scores are recorded for both SBDMCMC and Lasso algorithms. In Figure \ref{fig:f1score}, we observe that the proposed SBDMCMC method achieves much higher $F_1$-scores compared to LASSO method.

\section{Real Data Analysis}\label{sec:6}
Our method is applied on an influenza vaccination dataset. Influenza vaccinations are widely acknowledged as the primary method for protecting individuals against infections caused by influenza viruses. These vaccinations play an instrumental role in public health strategies worldwide, aiming to reduce the occurrence and impact of the seasonal flu. After receiving a vaccination, an individual will initiate an immune response that is impacted by their genetic factors. These genetic factors also affect susceptibility to influenza, indicating that our genes play a significant role in how we react to both vaccination and potential viral infections. We apply the proposed dynamic Ising model to build the gene network and explore the dependency structures.  This analysis could offer valuable insights into key genes that significantly impact the immune response to influenza vaccination. The results also reveal how the gene network is changing with respect to time.

We obtain the gene expression dataset (GSE48024) for influenza vaccination from the National Center for Biotechnology Information website
(\url{https://www.ncbi.nlm.nih.gov/geo/query/acc.cgi?acc=GSE48024}). The data set comprises longitudinal whole-blood gene expression profiles before and after seasonal influenza vaccination (TIV, trivalent inactivated vaccine). Gene expressions were measured using the Illumina HumanHT-12 v4.0 BeadChip microarray platform.
The dataset comprises 848 individuals, consisting of 119 healthy adult male volunteers and 128 healthy adult female volunteers. All individuals in the study have measurements of global transcript abundance in peripheral blood RNA samples at four time points, including one before vaccination and three after vaccination.

Following the methodology of \cite{franco2013integrative}, we choose to concentrate on 68 genes that display both a reaction at the transcriptional level to the vaccine and indications of genetic regulation. We dichotomize the measurements based on the mean for each gene. If a measurement exceeds the corresponding mean, it is set to 0; otherwise, it is set to 1. This process results in a binary dataset. 
In this dynamic Ising model, we incorporate a single covariate: the time points, which include day 0 (pre-influenza vaccination), day 1 (1 day post-influenza vaccination), day 3 (3 days post-influenza vaccination), and day 14 (14 days post-influenza vaccination). To capture dynamic immune responses, we apply the SBDMCMC approach to reveal the global gene structure dependent on the covariate time points.

\begin{figure}[ht]
\subfloat[Baseline graph structure  ]
{\includegraphics[width=0.45 \linewidth]{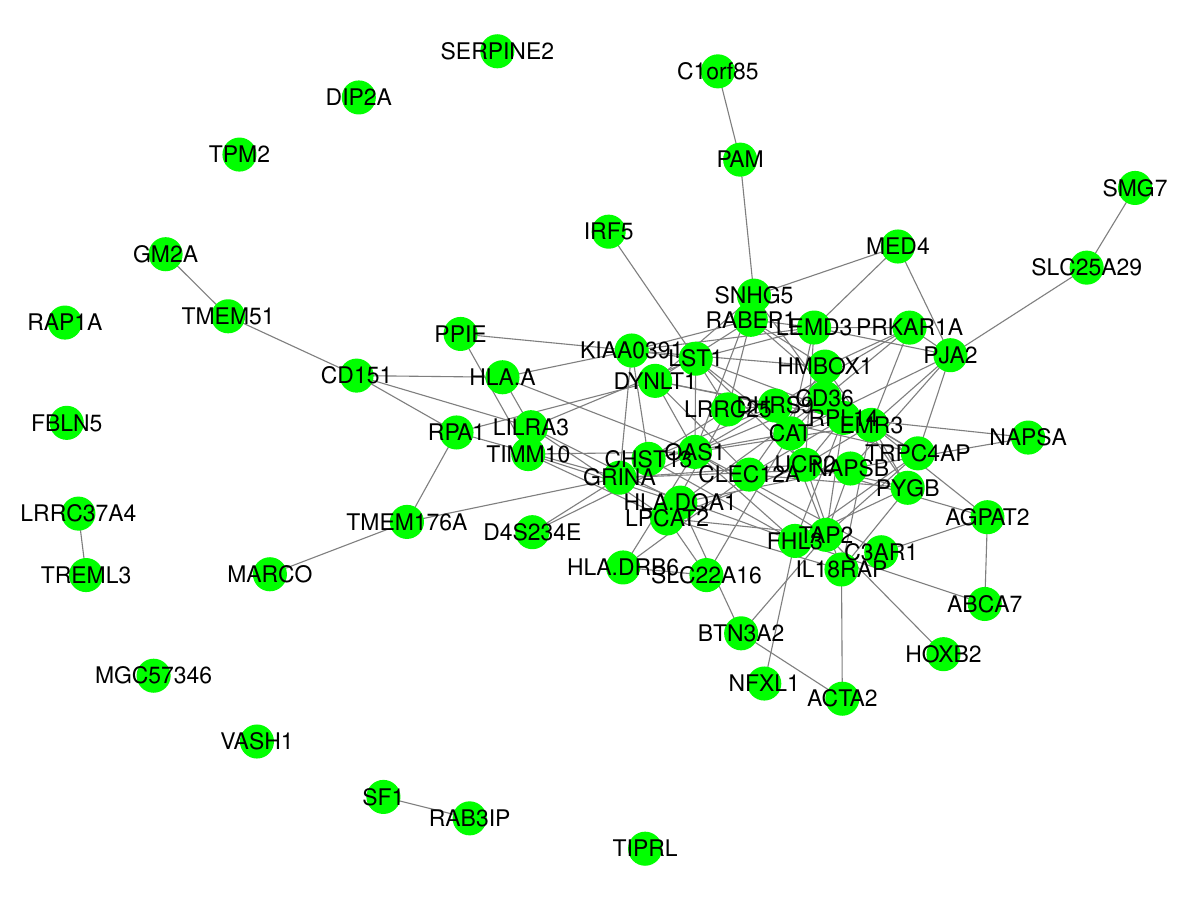} \label{fig:baseand}}
\subfloat[Slope graph structure]
{\includegraphics[width=0.45 \linewidth]{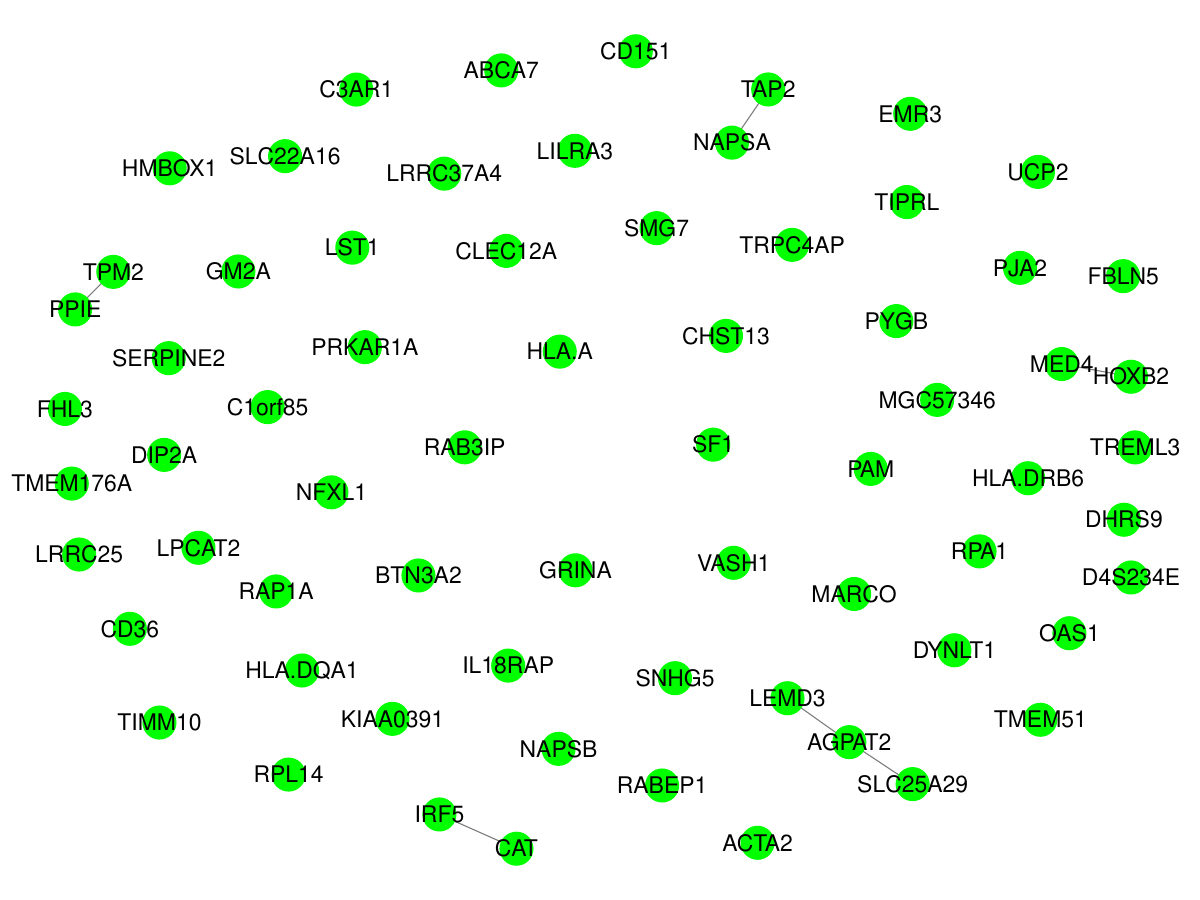} \label{fig:slopeand}}
\caption{Dynamic Ising model structure learning results based on ``AND" rule}
\label{fig:realdataand}
\end{figure}

\begin{figure}[ht]
\subfloat[Baseline graph structure  ]
{\includegraphics[width=0.45 \linewidth]{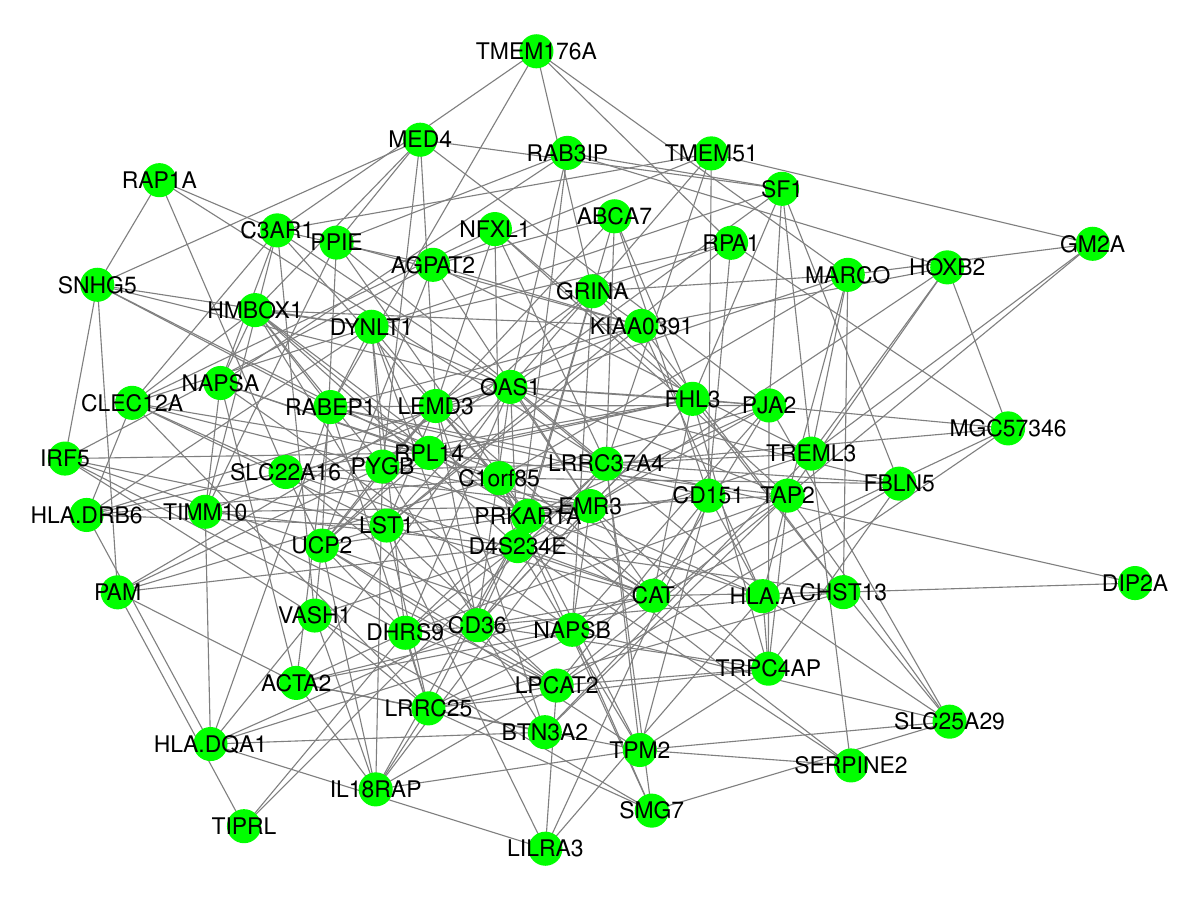} \label{fig:baseor}}
\subfloat[Slope graph structure]
{\includegraphics[width=0.45 \linewidth]{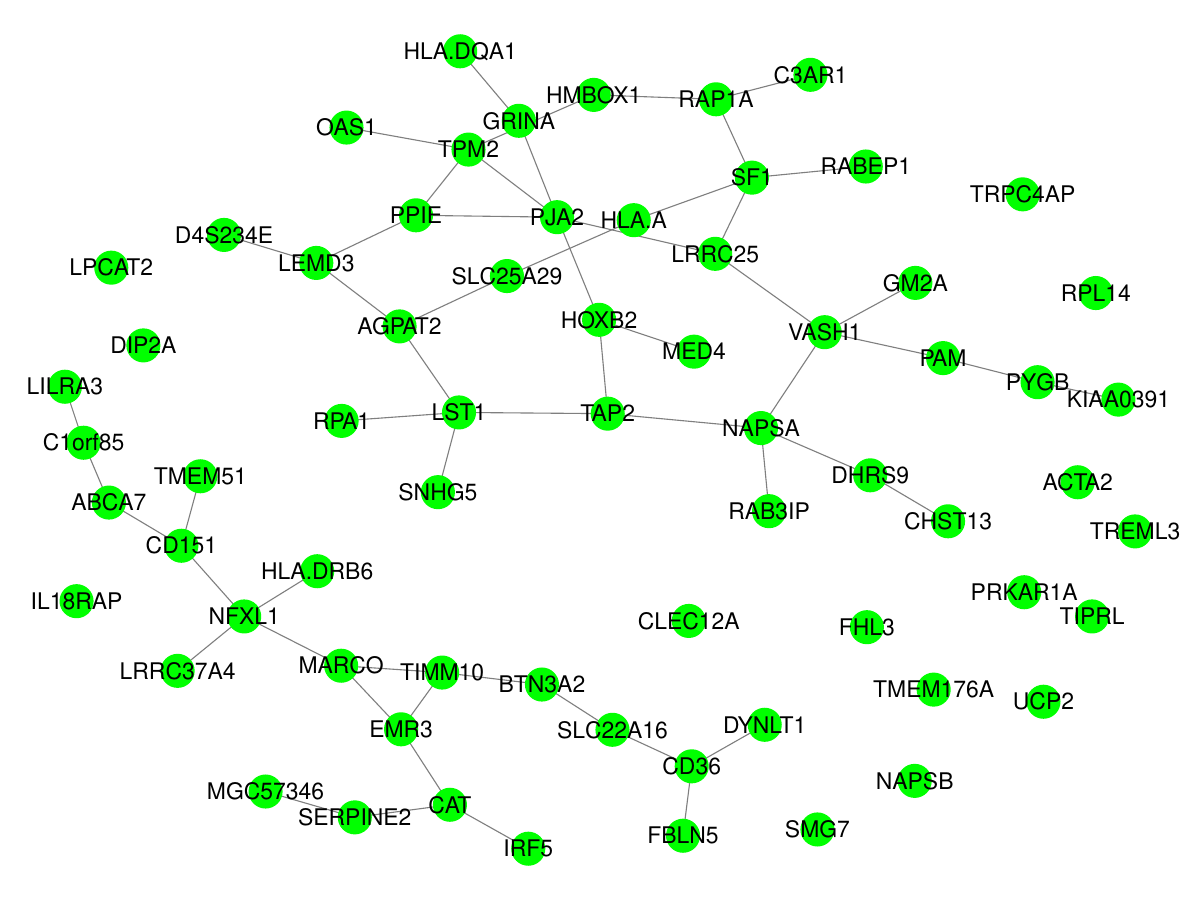} \label{fig:slopeor}}
\caption{Dynamic Ising model structure learning results based on ``OR" rule}
\label{fig:realdataor}
\end{figure}

As explained in Section \ref{sec:sbdmcmc}, SBDMCMC is a neighbourhood-based graphical model learning algorithm. Therefore, one needs to combine the neighbourhoods using either the ``AND" rule or the ``OR" rule to obtain the global model structure. The ``AND" rule favours a sparser structure, while the ``OR" rule uncovers more interactions among variables. In this analysis of influenza vaccination data, we present results from both approaches. The resulting gene network and sparsity patterns based on the ``AND" rule and ``OR" rule are illustrated in Figure \ref{fig:realdataand} and Figure \ref{fig:realdataor}, respectively. The estimated slope graph structure derived using the ``AND" rule results in a highly sparse network, as visually depicted in Figure \ref{fig:realdataand} (b). This graphical representation facilitates the identification of gene interactions that exhibit significant changes to vaccination across specified time points. Our analysis discovers several interactions among the following gene pairs: PPIE-TPM2, TAP2-NAPSA, MED4-HOXB2, IRF5-CAT, LEMD3-AGPAT2, and AGPAT2-SLC25A29. Notably, the genes TAP2 and NAPSA have been identified as exhibiting a transcriptional response to vaccination, demonstrating significant genotype effects on gene expression, as well as a correlation between transcriptional and antibody responses \citep{franco2013integrative}. These findings highlight specific gene-gene interactions that are influenced by vaccine administration and suggest potential targets for further biological investigation.

For any given time point, we can derive different gene interaction structures by combining the baseline graph and the slope graph using a linear form for that time point. This allows us to explore changes in gene relationships before (day 0) and after vaccination (day 1, day 3 and day 14). Furthermore, this analysis can reveal variations in gene interactions immediately after vaccination and in the days that follow. It provides insights into whether and how the vaccine influences the interplay between genes, which could have significant implications for understanding the underlying biological mechanisms and their potential impact on medical outcomes.

\section{Conclusion}

In this paper, we present a covariate-dependent discrete graphical model designed to capture dynamic networks with evolving dependence structures among discrete random variables. This model is well-suited for analyzing systems whose behaviour evolves over time and other covariates, with broad applications in fields such as healthcare, finance, and social networks. Our proposed framework models cell probabilities using log-linear parameters that vary with covariates, leveraging a novel parameterization based on a baseline graph and a slope graph, each defined by its own generating class. This structure provides a flexible and interpretable approach to modelling changing relationships between variables.

We perform efficient parameter estimation in high-dimensional settings using pseudo-likelihood estimation to infer the parameters of a dynamic Ising model with 100 variables and a single covariate. For model selection, we employ the SBDMCMC algorithm to learn the underlying model structure. We apply this approach to fit a covariate-dependent discrete graphical model to a gene expression dataset related to influenza vaccination, which includes longitudinal whole-blood gene expression profiles collected before and after vaccination. By integrating information from both baseline and slope graphs, we capture dynamic changes in gene interactions associated with the immune response to vaccination. Our results demonstrate the effectiveness of the proposed framework in uncovering meaningful, covariate-dependent relationships in high-dimensional discrete data, providing a powerful tool for understanding complex dynamic systems.


\bibliographystyle{apalike}
\bibliography{sample}

@article{castelo2006robust,
  title={A robust procedure for {G}aussian graphical model search from microarray data with $p$ larger than $n$.},
  author={Castelo, Robert and Roverato, Alberto},
  journal={Journal of Machine Learning Research},
  volume={7},
  number={12},
  pages={2621-2650},
  year={2006}
}

@article{roverato2002hyper,
  title={Hyper inverse Wishart distribution for non-decomposable graphs and its application to {B}ayesian inference for {G}aussian graphical models},
  author={Roverato, Alberto},
  journal={Scandinavian Journal of Statistics},
  volume={29},
  number={3},
  pages={391--411},
  year={2002},
  publisher={Wiley Online Library}
}

@article{kolodziejek2023discrete,
  title={Discrete parametric graphical models with Dirichlet type priors},
  author={Ko{\l}odziejek, Bartosz and Weso{\l}owski, Jacek and Zeng, Xiaolin},
  journal={arXiv preprint arXiv:2301.06058},
  year={2023}
}

@article{model_selection_2022,
  title = {Model Selection in the Space of {G}aussian Models Invariant by Symmetry},
  author = {Piotr Graczyk, Hideyuki Ishi, Bartosz Kołodziejek and Hélène Massam},
  journal = {Annals of Statistics},
  year = {2022},
  volume = {50},
  number = {3},
  pages = {1747--1774}
}

@article{niu2024covariate,
  title={Covariate-assisted {B}ayesian graph learning for heterogeneous data},
  author={Niu, Yabo and Ni, Yang and Pati, Debdeep and Mallick, Bani K.},
  journal={Journal of the American Statistical Association},
  volume={119},
  number={547},
  pages={1985--1999},
  year={2024},
  publisher={Taylor \& Francis}
}

@article{CHEN2025100984,
title = {A probabilistic modeling framework for genomic networks incorporating sample heterogeneity},
journal = {Cell Reports Methods},
volume = {5},
number = {2},
pages = {100984},
year = {2025},
author = {Liying Chen and Satwik Acharyya and Chunyu Luo and Yang Ni and Veerabhadran Baladandayuthapani},
}

@article{dobra2018loglinear,
	title = {Loglinear model selection and human mobility},
	volume = {12},
	number = {2},
    pages = {815--845},
	journal = {The Annals of Applied Statistics},
	author = {Dobra, Adrian and Mohammadi, Reza},
	year = {2018}
}

@article{mohammadi2015bayesian,
	title = {Bayesian structure learning in sparse {G}aussian graphical models},
	volume = {10},
	number = {1},
	urldate = {2025-07-16},
	journal = {Bayesian Analysis},
	author = {Abdolreza Mohammadi and Ernst C. Wit},
	pages = {109--138},
	year = {2015}
}

@article{wang2023scalable,
  title={The scalable birth-death {MCMC} algorithm for mixed graphical model learning with application to genomic data integration},
  author={Wang, Nanwei and Massam, H{\'e}l{\`e}ne and Gao, Xin and Briollais, Laurent},
  journal={The Annals of Applied Statistics},
  volume={17},
  number={3},
  pages={1958-1983},
  year={2023}
}

@article{guo2011joint,
  title={Joint estimation of multiple graphical models},
  author={Guo, Jian and Levina, Elizaveta and Michailidis, George and Zhu, Ji},
  journal={Biometrika},
  volume={98},
  number={1},
  pages={1--15},
  year={2011},
  publisher={Oxford University Press}
}

@article{lauritzendawid1993,
  title={Hyper Markov laws in the statistical analysis of decomposable graphical models},
  author={Dawid, Arnold P. and Lauritzen, Steffen},
  journal={The Annals of Statistics},
  pages={1272--1317},
  volume={21},
  number={3},
  year={1993},
  publisher={IMS}
}

@book{lauritzen1996gms,
  title={Graphical Models},
  author={Lauritzen, Steffen},
  year={1996},
  publisher={Oxford Science Publications}
}

@article{2012Bayes,
  title={Bayes factors and the geometry of discrete hierarchical loglinear models},
  author={ Letac, Gerard  and  Massam, H\'{e}l\`{e}ne},
  journal={The Annals of Statistics},
  volume={40},
  number={2},
  pages={861-890},
  year={2012},
}

@article{massamliudobra2009,
  title={A Conjugate Prior for Discrete Hierarchical Log-Linear Models},
  author={Hélène Massam and Jinnan Liu and Adrian Dobra},
  journal={The Annals of Statistics},
  pages={3431--3467},
  volume={37},
  number={6A},
  year={2009},
  publisher={JSTOR}
}

@article{zhang2023high,
  title={High-dimensional {G}aussian graphical regression models with covariates},
  author={Zhang, Jingfei and Li, Yi},
  journal={Journal of the American Statistical Association},
  volume={118},
  number={543},
  pages={2088--2100},
  year={2023},
  publisher={Taylor \& Francis}
}

@article{lin2017joint,
  title={On joint estimation of {G}aussian graphical models for spatial and temporal data},
  author={Lin, Zhixiang and Wang, Tao and Yang, Can and Zhao, Hongyu},
  journal={Biometrics},
  volume={73},
  number={3},
  pages={769--779},
  year={2017},
  publisher={Wiley Online Library}
}

@article{ni2022bayesian,
  title={Bayesian covariate-dependent {G}aussian graphical models with varying structure},
  author={Ni, Yang and Stingo, Francesco, C. and Baladandayuthapani, Veerabhadran},
  journal={Journal of Machine Learning Research},
  volume={23},
  number={242},
  pages={1--29},
  year={2022}
}

@article{liu2010graph,
  title={Graph-valued regression},
  author={Liu, Han and Chen, Xi and Wasserman, Larry and Lafferty, John},
  journal={Advances in Neural Information Processing Systems},
  volume={23},
  year={2010}
}

@article{franco2013integrative,
  title={Integrative genomic analysis of the human immune response to influenza vaccination},
  author={Franco, Luis M. and Bucasas, Kristine L. and Wells, Janet M. and Ni{\~n}o, Diane and Wang, Xueqing and Zapata, Gladys E. and Arden, Nancy and Renwick, Alexander and Yu, Peng and Quarles, John M. and others},
  journal={Elife},
  volume={2},
  pages={e00299},
  year={2013},
  publisher={eLife Sciences Publications, Ltd}
}

@book{wainwrightjordan2008,
  author={Wainwright, Martin J. and Jordan, Michael I.},
  title={Graphical Models, Exponential Families, and Variational Inference},
  year={2008},
  volume={},
  number={},
  pages={},
  publisher={Now Foundations and Trends},
  doi={10.1561/2200000001}}

@article{10.1214/14-BA931,
author = {Riten Mitra and Peter M{\"u}ller and Yuan Ji},
title = {{Bayesian graphical models for differential pathways}},
volume = {11},
journal = {Bayesian Analysis},
number = {1},
publisher = {International Society for Bayesian Analysis},
pages = {99--124},
year = {2016}
}

@article{peterson2015bayesian,
  title={Bayesian inference of multiple {G}aussian graphical models},
  author={Peterson, Christine and Stingo, Francesco C and Vannucci, Marina},
  journal={Journal of the American Statistical Association},
  volume={110},
  number={509},
  pages={159--174},
  year={2015},
  publisher={Taylor \& Francis}
}

@article{cheng2014sparse,
  title={A sparse {I}sing model with covariates},
  author={Cheng, Jie and Levina, Elizaveta and Wang, Pei and Zhu, Ji},
  journal={Biometrics},
  volume={70},
  number={4},
  pages={943--953},
  year={2014},
  publisher={Oxford University Press}
}

@article{Tan_2017,
   title={Bayesian inference for multiple {G}aussian graphical models with application to metabolic association networks},
   volume={11},
   number={4},
   pages={2222--2251},
   journal={The Annals of Applied Statistics},
   author={Tan, Linda S. L. and Jasra, Ajay and De Iorio, Maria and Ebbels, Timothy M. D.},
   year={2017}
  }

@article{RoachGao2023,
   title={Graphical Local Genetic Algorithm for High-Dimensional Log-Linear Models},
   volume={11},
   number={11},
   journal={Mathematics},
   author={Roach, Lyndsay and Gao, Xin},
   year={2023}}

@article{RoachMassamWangGao2025,
   title={Bayesian model selection consistency for high-dimensional discrete graphical models},
   volume={in press},
   journal={Bernoulli},
   author={Roach, Lyndsay and Massam, H\'{e}l\`{e}ne and Wang, Nanwei and Gao, Xin},
   year={2025}}

\end{document}